\documentclass[draft]{scrartcl}

\usepackage{url}
\newcommand{\email}[1]{\url{#1}}

\usepackage[utf8x]{inputenc}
\usepackage[T1]{fontenc}
\usepackage{xcolor}
\usepackage{latexsym} 
\usepackage{amsfonts} 
\usepackage{amssymb}
\usepackage{amsmath}
\usepackage{mathrsfs}
\usepackage{amsthm}
\usepackage{textcomp}
\usepackage{pifont}
\usepackage{stmaryrd}
\usepackage[mathscr]{eucal}
\usepackage{slashed} 
\usepackage[english]{babel}
\usepackage{scrpage2}
\usepackage{tikz}
\usetikzlibrary{arrows,decorations.pathmorphing,positioning,shapes,snakes,backgrounds,fit}
\usepackage[numbers 
]{natbib}   
\usepackage{tr}

\newtheorem{theorem}{Theorem}[section]

\newtheorem{lemma}[theorem]{Lemma}

\newtheorem{proposition}[theorem]{Proposition}
\theoremstyle{definition}      
\newtheorem{definition}[theorem]{Definition}

\newtheorem{example}[theorem]{Example}

\deffootnote{1em}{1em}{\textsuperscript{\thefootnotemark\ }}

\begin{document}

\title{Dependency Pairs and Polynomial Path Orders%
\thanks{This research is partially supported by FWF (Austrian Science Fund) projects P20133.}}
\author{
Martin Avanzini
and
Georg Moser\\
\{martin.avanzini,georg.moser\}@uibk.ac.at
}
\date{March 2009}

\maketitle

\begin{abstract}
We show how polynomial path orders can be employed efficiently in 
conjunction with weak innermost dependency pairs to automatically certify 
polynomial runtime complexity of term rewrite systems and
the polytime computability of the functions computed.
The established techniques have been implemented and we provide ample
experimental data to assess the new method.
\end{abstract}

\tableofcontents
\newpage

\section{Introduction} \label{s:intro}

In order to measure the complexity of a (terminating) term rewrite system (TRS for short) 
it is natural to look at the maximal length of derivation sequences---the \emph{derivation length}---as
suggested by Hofbauer and Lautemann in~\cite{HL89}. More precisely, 
the \emph{runtime complexity function} with respect to a (finite and terminating) TRS $\RS$ 
relates the maximal derivation length to the size of the initial term, whenever
the set of initial terms is restricted to constructor based terms, also 
called \emph{basic} terms. The restriction to basic terms allows us to
accurately express the complexity of a program through the runtime complexity of TRSs.
In this paper we study and combine recent efforts for the \emph{automatic} analysis of runtime complexities of TRSs.
In~\cite{AvanziniMoser:2008} we introduced a restriction of the multiset path order,
called \emph{polynomial path order} (\emph{\POPSTAR} for short) 
that induces polynomial runtime complexity if restricted to innermost rewriting.
The definition of \POPSTAR\ employs the idea of \emph{tiered recursion}~\cite{Simmons:1988}.
Syntactically this amounts to a  separation of arguments into \emph{normal} and \emph{safe} arguments, 
cf.~\cite{BellantoniCook:1992}.
Furthermore, Hirokawa and the second author introduced a variant of dependency pairs, dubbed
\emph{weak dependency pairs}, that makes the dependency pair method applicable in the context
of complexity analysis, cf.~\cite{HirokawaMoser:2008,HirokawaMoser2:2008}. 

We show how weak innermost dependency pairs can be successfully applied
in conjunction with \POPSTAR. The following example (see \cite{FGMSTZ:2007}) motivates this study.
Consider the TRS $\RSbin$ encoding the function $\lambda x. \roundabove{\log(x+1)}$ 
for natural numbers given as tally sequences:
\begin{alignat*}{4}
  1\colon &&\halfs(\mN) & \to \mN &  
  4\colon && \bits(\mN)& \to \mN 
  \\
  2\colon &&\halfs(\ms(\mN))& \to \mN &  
  5\colon &&\bits(\ms(\mN))& \to \ms(\mN)
  \\
  3\colon && \halfs(\ms(\ms(x)))& \to \ms(\halfs(x)) &  \hspace{2ex}
  6\colon &&\bits(\ms(\ms(x))) & \to
  \ms(\bits(\ms(\halfs(x))))  
\end{alignat*}
It is easy to see that the TRS $\RSbin$ is not compatible with \POPSTAR, 
even if we allow quasi-precedences, see Section~\ref{s:pop}. 
On the other hand, employing (weak innermost) dependency pairs, argument filtering, and
the usable rules criteria in conjunction with $\POPSTAR$, polynomial innermost runtime complexity of $\RSbin$ can be shown 
fully automatically. 

The combination of dependency pairs and polynomial path orders, while conceptually quite clear,
turns out to be technical involved. One of the first obstacles one encounters is that 
the pair $(\geqpop,\gpop)$ cannot be used as a reduction pair in the spirit of~\cite{HirokawaMoser:2008},
as $\geqpop$ fails to be closed under contexts. Conclusively, we start 
from scratch and study polynomial path orders in the context of \emph{relative rewriting}~\cite{Geser:1990}.
Based on this study an incorporation of argument filterings becomes possible so that we can employ
the pair $(\geqpopp,\gpopp)$ in conjunction with dependency pairs successfully. 
Here, $\gpopp$ refers to the order obtained by combining $\gpop$ with
the argument filtering $\pi$
as expected, and $\geqpopp$ denotes the extension of $\gpopp$ by
term equivalence, preserving the separation of safe and
normal argument positions. 
Note that for polynomial path orders, the integration of argument filterings is not only
non-trivial, but indeed a challenging task. This is mainly due to the embodiment of tiered
recursion in \POPSTAR.
Thus we establish a combination of two syntactic techniques in complexity analysis. 
The experimental evidence given below indicates the power and in particular the efficiency 
of the provided results. 

Our next contribution is concerned with \emph{implicit complexity theory}, 
see for example~\cite{BMM:2009:tcs}. A careful analyis of our main result shows 
that polynomial path orders in conjunction with (weak innermost) dependency pairs even induce polytime 
computability of the functions defined by the TRS studied. This result
fits well with recent results by Marion and P{\'e}choux on the use of restricted forms 
of the dependency pair method to charcterise complexity classes like $\ptime$ or 
$\pspace$, cf.~\cite{MP08b}. 
Note that both results allow to conclude, based on different restrictions, 
polytime computability of the functions defined by constructor TRSs, 
whose termination can be shown by the dependency pair method. Note
that the results in~\cite{MP08b} also capture programs admitting infeasible
runtime complexities but define functions that are computable in polytime if suitable (and
non-trivial) program transformations are used. Such programs are outside the scope
of our results. Thus it seems that our results more directly assess the complexity
of the given programs. 
Note that our tool provides (for the first time) a fully automatic application of the dependency
pair method in the context of implicit complexity theory.%
\footnote{In this context it is perhaps interesting to note that for a variant of the TRS $\RSbin$,
studied in~\cite{MP08b}, our tool verifies polytime computability fully
automatically.
See also~\cite{AvanziniMoserSchnabl:2008}
for the description of a small tool that implements related characterisations
of of the class of polynomial time computable functions.}

\medskip
The rest of the paper is organised as follows. In Section~\ref{s:seq} we 
present basic notions and recall (briefly) the \emph{path order for \fptime} from \cite{AraiMoser:2005}. 
We then briefly recall dependency pairs in the context of complexity analysis from 
\cite{HirokawaMoser:2008,HirokawaMoser2:2008}, cf.~Section~\ref{s:dp}. 
In Section~\ref{s:pop} we present polynomial path orders over quasi-precedences. Our main results
are presented in Section \ref{s:dppop}. We continue with
experimental results in Section \ref{s:exps}, and conclude in Section~\ref{s:concl}.

\section{The Polynomial Path Order on Sequences} 
\label{s:seq}

We assume familiarity with the basics of term rewriting, see~\cite{BN98,TeReSe}.
Let $\VS$ denote a countably infinite set of variables and $\FS$ a 
signature, containing at least one constant. 
The set of terms over $\FS$ and $\VS$ is denoted as $\TERMS$
and the set of ground terms as $\TA(\FS)$.
We write $\Fun(t)$ and $\Var(t)$ for the set of function symbols and
variables appearing in $t$, respectively. 
The root symbol $\rt(t)$ of a term $t$ is defined as usual and 
the (proper) subterm relation is denoted as $\subterm$ ($\subtermstrict$). 
We write $\subtermat{s}{p}$ for the \emph{subterm} of $s$ at position $p$.
The \emph{size} $\size{t}$ of a term $t$ is defined as usual and
the \emph{width} of $t$ is defined as $\width(t) \defi \max\set{n,\map{\width}{t}}$ if 
$t = f(\seq{t})$ and $n > 0$ or $\width(t) = 1$ else.
Let $\qp$ be a preorder on the signature $\FS$, called \emph{quasi-precedence}
or simply \emph{precedence}. Based on $\qp$ we 
define an equivalence $\eqi$ on terms:
$s \eqi t$ if either (i) $s=t$ or (ii) $s = f(\seq{s})$, $t =
g(\seq{t})$,  $f \eqi g$ and there exists a permutation
$\pi$ such that $s_i \eqi t_{\pi(i)}$. For a preorder $\succsim$, we 
use $\mextension{\succsim}$ for the multiset
extension of $\succsim$, which is again a preorder.
The proper order (equivalence) induced by $\mextension{\succsim}$
is written as $\mextension{\succ}$ ($\mextension{\approx}$).

A \emph{term rewrite system} (\emph{TRS} for short) $\RS$ over
$\TERMS$ is a \emph{finite} set of rewrite
rules $l \to r$, such that $l \notin \VS$ and $\Var(l) \supseteq \Var(r)$.
We write $\rew$ ($\irew$) for the induced (innermost) rewrite
relation. The set of defined function symbols is denoted as $\DS$, while the constructor
symbols are collected in $\CS$, clearly $\FS = \DS \cup \CS$.
We use $\NF(\RS)$ to denote the set of normal forms of $\RS$ and
set $\Val \defi \TA(\CS,\VS)$, the elements of $\Val$ are called
\emph{values}. A TRS is called \emph{completely defined} if normal forms
coincide with values.
We define $\Tb \defi \{f(\seq{v}) \mid f \in \DS 
\text{ and } v_i \in \Val\}$ as the set of \emph{basic terms}.
A TRS $\RS$ is a \emph{constructor TRS} if $l \in \Tb$ for 
all ${l \to r} \in \RS$.
Let $\QS$ denote a TRS. 
The \emph{generalised restricted rewrite relation $\qrew[\RS]{\QS}$}
is the restriction of $\rew$ where all arguments of the redex are in
normal form with respect to the TRS $\QS$ (see~\cite{Thiemann:2007}). 
\label{d:qrew}
We define the (innermost) relative rewriting relation (denoted as $\irrew{\RS}{\RSS}$) as
follows:
\begin{equation*}
{\irrew{\RS}{\RSS}} \defi
  {{\qrss[\RSS]{\RS \cup \RSS}} 
    \cdot {\qrew[\RS]{\RS \cup \RSS}} 
    \cdot {\qrss[\RSS]{\RS \cup \RSS}}} \tpkt
\end{equation*}
Similarly, we set 
  ${\irrewt{\RS}{\RSS}} \defi
  {{\qrss[\RSS]{\RS \cup \RSS}} 
    \cdot {\qrewt[\RS]{\RS \cup \RSS}} 
    \cdot {\qrss[\RSS]{\RS \cup \RSS}}}$,
to define an \emph{(innermost) relative root-step}.

A \emph{polynomial interpretation} is a well-founded and monotone algebra $(\AS,>)$ with carrier 
$\NAT$ such that $>$ is the usual order on natural numbers and all interpretation 
functions $f_\AS$ are polynomials. Let $\alpha\colon \VS \to \AS$ denote an 
\emph{assignment}, then we write $\interpret[\AS]{\alpha}{t}$ for the evaluation
of term $t$ with respect to $\AS$ and $\alpha$. A polynomial interpretation
is called a \emph{strongly linear interpretation} (\emph{SLI} for short)
if all function symbols are interpreted by 
\emph{weight functions} $f_\AS(\seq{x}) = \sum_{i=1}^n x_i + c$ with
$c \in \NAT$.
The \emph{derivation length} of a terminating term $s$ with respect to
$\to$ is defined as $\dl(s,\to) \defi \max\set{ n \mid \exists t. \; s \to^n t }$,
where $\to^n$ denotes the $n$-fold application of $\to$. The 
\emph{innermost runtime complexity function} $\rcRi$ with respect to
a TRS $\RS$ is defined as 
$\rcRi(n) \defi \max\{ \dl(t, \irew) \mid \text{$t \in \Tb$ and $\size{t} \leqslant n$} \}$.
If no confusion can arise $\rcRi$ is simply called \emph{runtime complexity function}.

Below we recall the bare essentials of the polynomial path order
$\gpopv$ on sequences (\POP~for short) as put forward in~\cite{AraiMoser:2005}.
We kindly refer the reader to~\cite{AraiMoser:2005,AvanziniMoser:2008} for motivation and examples.
We recall the definition of \emph{finite approximations $\gpopv_k^l$} of $\gpopv$.  The latter 
is conceived as the \emph{limit} of these approximations.
The domain of this order are so called \emph{sequences}
$\SE(\FS,\VS) \defi \TA(\FS\cup\{\circ\},\VS)$. Here $\FS$ is a finite
signature and $\circ \not \in \FS$ a fresh variadic function symbol, 
used to form sequences.
We denote sequences $\circ(\seq{s})$ by $[\sexpr{s}]$ and write
$a \cons [\sexpr{b}]$ for the sequence $[a~\sexpr{b}]$.

Let $\qp$ denote a precedence. The order $\gpopv_k^l$ is based on an auxiliary order $\gppv_k^l$
(and the equivalence $\eqi$ on terms defined above).
Below we set ${\geqppv_k^l} \defi {\gppv_k^l} \cup {\eqi}$.
We write $\multiset{t_1,\dots,t_n}$ to denote multisets 
and $\uplus$ for the multiset sum.

\begin{definition} \label{d:approx:sq}
Let $k,l \geqslant 1$.
The order $\gppv_{k}^{l}$
induced by $\qp$ is inductively defined as follows:
$s \gppv_{k}^{l} t$  for $s=f(\seq{s})$ or $s=[\sexpr{s}]$ if either
\begin{enumerate}
\item $s_i~\geqppv_{k}^{l}~t$ for some $i \in \set{1,\dots,n}$, or
\item \label{en:approx:sq:2}
$s=f(\seq{s})$, $t = g(\seq[m]{t})$ with $f \succ g$ or $t=[\sexpr[m]{t}]$, 
  $s \gppv_{k}^{l-1} t_j$ for all $j \in \set{1,\dots,m}$, and $m < k+\width(s)$,
\item $s=[\sexpr{s}]$, $t = [\sexpr[m]{t}]$ and the following properties hold:
  \begin{itemize}
  \item $\multiset{\seq[m]{t}} = N_1 \uplus \cdots \uplus N_n$ 
    for some multisets $N_1,\dots, N_n$, and  
  \item there exists $i \in \set{1,\dots,n}$ such that $\multiset{s_i} \not\eqi^\mul N_i$, and
  \item for all $1 \leqslant i \leqslant n$ such that $\multiset{s_i} \not\eqi^\mul N_i$ 
    we have $s_i \gppv_k^l r$ for all $r \in N_i$, and $m < k + \width(s)$.
  \end{itemize}
\end{enumerate}
\end{definition}
\begin{definition} \label{d:approx}
Let $k,l \geqslant 1$.
The \emph{approximation $\gpopv_{k}^{l}$ of the polynomial path order on sequences} 
induced by $\qp$ is inductively defined as follows:
$s \gpopv_{k}^{l} t$ for $s=f(\seq{s})$ or $s=[\sexpr{s}]$ if either
$s \gppv_k^l t$ or 
\begin{enumerate}
\item $s_i \geqpopv_k^l t$ for some $i \in \set{1,\dots,n}$,
\item\label{d:approx:ii} $s = f(\seq{s})$, $t = [\sexpr[m]{t}]$, and the following
  properties hold:
  \begin{itemize}
  \item $s \gpopv_k^{l-1} t_{j_0}$ for some $j_0  \in \set{1,\dots,m}$, 
  \item $s \gppv_k^{l-1} t_j$ for all $j \neq j_0$, and $m < k + \width(s)$,
  \end{itemize}
\item\label{d:approx:iii} $s = f(\seq{s})$, $t = g(\seq[m]{t})$, $f \sim g$ and $[\sexpr{s}] \gpopv_k^l [\sexpr[m]{t}]$, or
\item\label{d:approx:iv} $s=[\sexpr{s}]$, $t = [\sexpr[m]{t}]$ and the following properties hold:
  \begin{itemize}
  \item $\multiset{\seq[m]{t}} = N_1 \uplus \cdots \uplus N_n$ for
    some multisets $N_1,\dots, N_n$, and 
  \item there exists $i \in \set{1,\dots,n}$ such that $\multiset{s_i} \not\eqi^\mul N_i$, and
  \item for all $1 \leqslant i \leqslant n$ such that $\multiset{s_i}\not\eqi^\mul N_i$ we have 
    $s_i \gpopv_k^l r$ for all $r \in N_i$, and $m < k + \width(s)$.
  \end{itemize}
\end{enumerate}
\end{definition}

Above we set ${\geqpopv_k^l} \defi {\gpopv_k^l} \cup {\eqi}$ and 
abbreviate $\gpopv_k^k$ as $\gpopv_k$ in the following. Note that the
empty sequence is minimal with respect to both orders.
It is easy to see that for $k \leqslant l$, we have ${\gppv_{k}} \subseteq {\gppv_{l}}$ and 
${\gpopv_{k}} \subseteq {\gpopv_{l}}$. Note that 
$s \gpopv_k t$ implies that  $\width(t) < \width(s) + k$. 
For a fixed approximation $\gpopv_k$, we define the length of its
longest decent as follows:
$\Slow(t) \defi \max\set{n \mid t = t_0 \gpopv_k t_1 \gpopv_k \dots \gpopv_k t_n}$.
The following proposition is a reformulation of~\cite[Lemma 6]{AraiMoser:2005}.
\begin{proposition}
  \label{p:popseq}
  Let $ k \in \NAT$.
  There exists a polynomial interpretation
  $\Ipop$ such that $\Slow(t) \leqslant \interpret[\Ipop]{\alpha}{t}$ 
  for all assignments $\ofdom{\alpha}{\VS \to \NAT}$. 
  As a consequence, for all terms $f(\seq{t})$
  with $\interpret[\Ipop]{\alpha}{t_i} = \bigO(\size{t_i})$, 
  $\Slow(f(\seq{t}))$ is bounded by a polynomial $p$ in the size of $t$, where
  $p$ depends on $k$ only.
\end{proposition}

Observe that the polynomial interpretation $\Ipop$ as employed in
the proposition fulfils: $\circ_\Ipop(\seq{m}) = \sum_{i=1}^n m_i + n$.
In particular, we have $\interpret[\Ipop]{\alpha}{[]} = 0$.

\section{Complexity Analysis Based on the Dependency Pair Method} \label{s:dp}

In this section, we briefly recall the central definitions and results
established in \cite{HirokawaMoser:2008,HirokawaMoser2:2008}.
We kindly refer the reader to
\cite{HirokawaMoser:2008,HirokawaMoser2:2008}
for further examples and underlying intuitions.
Let $\XS$ be a set of symbols. 
We write $\Ctx{\seq{t}}_\XS$ to denote $C[\seq{t}]$, whenever $\rt(t_i) \in \XS$ 
for all $i \in \set{1,\dots,n}$
and $C$ is a $n$-hole context containing no symbols from $\XS$.
We set $\mrk{\DS} \defi \DS \cup \set{\mrk{f} \mid f \in \DS}$ with 
each $\mrk{f}$ a fresh function symbol. Further, for 
$t = f(\seq{t})$ with $f \in \DS$, we set $\mrk{t} \defi \mrk{f}(\seq{t})$.
\begin{definition}
  Let $\RS$ be a TRS. If $l \to r \in \RS$ and $r = \Ctx{\seq{u}}_{\DS}$ then 
  $\mrk{l} \to \COM(\mrk{u_1},\ldots,\mrk{u_n})$ is called
  a \emph{weak innermost dependency pair} of $\RS$. 
  Here $\COM(t) = t$ and $\COM(\seq{t}) = \mc(t_1,\ldots,t_n)$, $n
  \not= 1$, for a fresh
  constructor symbol $\mc$, the \emph{compound symbol}.
  The set of all weak innermost dependency pairs is denoted by $\WIDP(\RS)$.
\end{definition}

\begin{example}\label{ex:bits:widp}
  Reconsider the example $\RSbits$ from the introduction.
  The set  of weak innermost dependency pairs $\WIDP(\RSbits)$ is given by
  \begin{alignat*}{4}
    7\colon &\;&\mrk{\halfs}(\mN)& \to \mathsf{c_1} & 
    10\colon &\;& \mrk{\bits}(\mN)& \to \mathsf{c_3} 
    \\
    8\colon &&\mrk{\halfs}(\ms(\mN))& \to \mathsf{c_2} & 
    11\colon && \mrk{\bits}(\ms(\mN))& \to \mathsf{c_4} 
    \\
    9\colon &&\mrk{\halfs}(\ms(\ms(x)))& \to \mrk{\halfs}(x) & \hspace{3ex}
    12\colon && \mrk{\bits}(\ms(\ms(x)))& \to \mrk{\bits}(\ms(\halfs(x))) 
  \end{alignat*}
\end{example}

We write $f \depends g$ if there exists a rewrite rule
$l \to r \in \RS$ such that $f = \rt(l)$ and $g$ is a defined 
symbol in $\Fun(r)$. 
For a set $\GS$ of defined symbols 
we denote by $\RS{\restriction}\GS$ the set of
rewrite rules $l \to r \in \RS$ with $\rt(l) \in \GS$. 
The set $\URs(t)$ of usable rules of a term $t$ is defined as
$\RS{\restriction}\set{g \mid \text{$f \depends^* g$ for some $f \in \Fun(t)$} }$.
Finally, we define $\URs(\PS) = \bigcup_{l \to r \in \PS} \URs(r)$.

\begin{example}[Example \ref{ex:bits:widp} continued]
  \label{ex:bits:us}
The usable rules of $\WIDP(\RSbits)$ consist of the following rules:
$1\colon \halfs(\mN) \to \mN$, $2\colon \halfs(\ms(\mN)) \to \mN$, and $3\colon \halfs(\ms(\ms(x))) \to  \halfs(x)$.
\end{example}
  
The following proposition allows the analysis of 
the (innermost) runtime complexity through the study of (innermost) relative rewriting,
see~\cite{HirokawaMoser:2008} for the proof.
\begin{proposition}
\label{p:usable}
\label{p:dp}
Let $\RS$ be a TRS, let $t$ be a basic terminating term, 
and let $\PS = \WIDP(\RS)$. 
Then $\dl(t, \irew) \leqslant \dl(t^{\sharp},\irew[\US(\PS)\,\cup \, \PS])$.
Moreover, if $\PS$ is non-duplicating and ${\URs(\PS)} \subseteq {>_\AS}$ for some SLI $\AS$.
Then there exist constants $K,L \geqslant 0$ (depending on $\PS$ and $\AS$ only) such that
  $\dl(t,\irew) \leqslant 
    K \cdot \dl(\mrk{t},\irrew{\PS}{\URs(\PS)}) + L \cdot \size{\mrk{t}}$.
\end{proposition}

This approach admits also an integration of \emph{dependency graphs} \cite{ArtsGiesl:2000}
in the context of complexity analysis. 
The nodes of the \emph{weak innermost dependency graph} $\WIDG(\RS)$
are the elements of $\PS$ and there is an arrow 
from $s \to t$ to $u \to v$ if there exist a context $C$ and substitutions $\sigma$, $\tau$
such that $t\sigma \irss[\RS] C[u\tau]$.
Let $\GS = \WIDG(\RS)$; a \emph{strongly connected component} (\emph{SCC} for short) 
in $\GS$ is a maximal \emph{strongly connected subgraph}.
We write $\PG{\GS}$ for the \emph{congruence graph}, where 
$\equiv$ is the equivalence relation induced by SCCs.
\begin{example}[Example \ref{ex:bits:widp} continued] 
  \label{ex:bits:dg}
  $\GS = \WIDG(\RSbits)$ consists of the nodes (7)--(12) as mentioned 
  in Example \ref{ex:bits:widp} and has the following shape:
  \begin{center}
    \begin{tikzpicture}[node distance=5mm]
      \node(7) {7} ;
      \node(9) [right=of 7] {9} ;
      \node(8) [right=of 9] {8} ;
      \node(10) [right=of 8] {10} ;
      \node(12) [right=of 10] {12} ;
      \node(11) [right=of 12] {11} ;
      
      \draw[->] (9) edge [loop above] (9);
      \draw[->] (9) edge (7);
      \draw[->] (9) edge (8);
      \draw[->] (12) edge [loop above] (12);
      \draw[->] (12) edge (11);
    \end{tikzpicture}
  \end{center}
The only non-trivial SCCs in $\GS$ are $\{9\}$ and $\{12\}$. Hence $\PG{\GS}$
consists of the nodes
$\equivclass[\equiv]{7}$--$\equivclass[\equiv]{12}$, and edges
$(\equivclass[\equiv]{a},\equivclass[\equiv]{b})$
for edges $(a,b)$ in $\GS$. Here 
$\equivclass[\equiv]{a}$ denotes the equivalence class of $a$.
\end{example}

We set $\LL(t) \defi \max\set{ \dl(t, \irrew{\PS_m}{\RSS}) \mid 
\text{$(\PS_1,\ldots,\PS_m)$ a path in $\PG{\GS}$, $\PS_1 \in \Src$}}$,
where $\Src$ denote the set o
f source nodes from $\PG{\GS}$ and
$\RSS = \PS_1 \cup \cdots \cup \PS_{m-1} \cup \URs(\PS_1 \cup \cdots \cup \PS_m)$.  
The proposition allows the use of different techniques
to analyse polynomial runtime complexity on separate paths, 
cf.~\cite{HirokawaMoser2:2008}.
\begin{proposition} 
\label{p:dg}
Let $\RS$, $\PS$, and $t$ be as above. 
Then there exists a polynomial $p$ (depending only on $\RS$) 
such that $\dl(\mrk{t},\irrew{\PS}{\URs(\PS)}) \leqslant p(\LL(\mrk{t}))$.
\end{proposition}

\section{The Polynomial Path Order over Quasi-Precedences} \label{s:pop}

In this section, we briefly recall the central definitions and results
established in \cite{AvanziniMoser:2008,AvanziniMoserSchnabl:2008} 
on the \emph{polynomial path order}. We employ the variant
of \POPSTAR\ based on quasi-precendences, cf.~\cite{AvanziniMoserSchnabl:2008}.

As mentioned in the introduction, \POPSTAR\ relies on tiered recursion, which
is captured by the notion of \emph{safe mapping}.
A \emph{safe mapping} $\safe$ is a function that associates 
with every $n$-ary function symbol $f$ the set of \emph{safe argument positions}.
If $f \in \DS$ then $\safe(f) \subseteq \set{1,\dots,n}$, for 
$f \in \CS$ we fix $\safe(f) = \set{1,\dots,n}$.
The argument positions not included in $\safe(f)$ are 
called \emph{normal} and denoted by $\normal(f)$.
We extend $\safe$ to terms $t \not \in \VS$ as follows: we define 
$\safe(f(\seq{t})) \defi \set{t_{i_1},\dots,t_{i_p}}$ where 
$\safe(f) = \set{\seq[p]{i}}$, likewise we define
$\normal(f(\seq{t})) \defi \set{t_{j_1},\dots,t_{j_q}}$ where $\normal(f) = \set{\seq[q]{j}}$.
Not every precedence is suitable for $\gpop$, in particular we need
to assert that constructors are minimal.

We say that a precedence $\qp$ is \emph{admissible} for \POPSTAR\
if the following is satisfied: 
(i) $f \succ g$ with $g \in \DS$ implies $f \in \DS$, and 
(ii) if $f \ep g$ then $f \in \DS$ if and only if $g \in \DS$. 
In the sequel we assume any precedence is admissible.
We extend the equivalence $\eqi$ to the
context of safe mapping: $s \eqis t$, if (i) $s = t$, or 
(ii) $s = f(\seq{s})$, $t = g(\seq{t})$, $f \ep g$ and there exists
a permutation $\pi$ so that $s_i \eqis t_{\pi(i)}$, where $i \in
\safe(f)$ if and only if $\pi(i) \in \safe(g)$ for all $i \in \set{1,\dots,n}$.
Similar to \POP, the definition of the polynomial path order $\gpop$ makes use of an
auxiliary order $\gsq$. 
\begin{definition} \label{d:pop2}
  The auxiliary order $\gsq$ induced by $\qp$ and $\safe$ is inductively defined as follows:
  $s = f(\seq{s}) \gsq t$ if either
\begin{enumerate}
\item\label{d:pop2:i} $s_i \geqsq t$ for some $i \in \set{1,\dots,n}$,
  and if $f \in \DS$ then $i \in \normal(f)$, or
\smallskip
\item\label{d:pop2:ii} $t = g(\seq[m]{t})$, $f \succ g$, $f \in \DS$
  and 
   $s \gsq t_j$ for all $j \in \set{1,\dots,m}$. 
\end{enumerate}
\end{definition}
\begin{definition} \label{d:pop3}
  The \emph{polynomial path order} $\gpop$  
  induced by $\qp$ and $\safe$ is inductively defined as follows:
  $s = f(\seq{s}) \gpop t$ if either
  $s \gsq t$ or
  \begin{enumerate}
  \item\label{d:pop3:i} $s_i \geqpop t$ for some $i \in \set{1,\dots,n}$, or
  \item\label{d:pop3:ii} $t = g(\seq[m]{t})$, $f \succ g$, $f \in
    \DS$, and  
    \begin{itemize}
    \item $s \gpop t_{j_0}$ for some $j_0 \in \safe(g)$, and
    \item for all $j \neq j_0$ either $s \gsq t_j$, or $s \rhd t_j$
      and $j \in \safe(g)$, or 
    \end{itemize}
  \item \label{d:pop3:iii} $t = g(\seq[m]{t})$, $f \ep g$,
    $\normal(s) \gpopmul \normal(t)$ and $\safe(s)\geqpopmul\safe(t)$.
  \end{enumerate}
\end{definition}
Above we set ${\geqsq} \defi {\gsq} \cup {\eqis}$ and 
${\geqpop} \defi {\gpop} \cup {\eqis}$ below. 
Here $\gpopmul$ and $\geqpopmul$ refer to the strict 
and weak multiset extension of $\geqpop$ respectively.

The intuition of $\gsq$ is to deny any recursive call, whereas
$\gpop$ allows predicative recursion:
by the restrictions imposed by $\safe$, recursion needs to be 
performed on normal arguments, while a recursively computed result 
must only be used in a safe argument position, compare~\cite{BellantoniCook:1992}.
Note that the alternative $s \rhd t_j$ for $j \in
\safe(g)$ in Definition \ref{d:pop3}(\ref{d:pop3:ii}) guarantees
that $\POPSTAR$ characterises the class of polytime computable
functions, cf.~\cite{AvanziniMoser:2008}. 
The proof of the next theorem follows the pattern of the proof 
of main theorem in~\cite{AvanziniMoser:2008}, but the
result is stronger due to the extension to quasi-precedences.
\begin{theorem} \label{t:popstar}
  Let $\RS$ be a constructor TRS. 
  If $\RS$ is compatible with $\gpop$, i.e., 
  ${\RS} \subseteq {\gpop}$, then the innermost runtime complexity $\rcRi$ induced
  is polynomially bounded.
\end{theorem}
 
Note that Theorem~\ref{t:popstar} is too weak to handle the TRS $\RSbits$
as the (necessary) restriction to an admissible precedence is too strong.
To rectify this, we suit \POPSTAR\ so that it can be used in conjunction
with weak (innermost) dependency pairs. 

An argument filtering (for a signature $\FS$) is a
mapping $\pi$ that assigns to every $n$-ary function symbol $f \in \FS$
an argument position $i \in \{ 1, \dots, n \}$ or a (possibly empty)
list $\{ \seq[m]{i} \}$ of argument positions with
$1 \leqslant i_1 < \cdots < i_m \leqslant n$.
The signature $\FSpi$ consists of all function symbols $f$ such that
$\pi(f)$ is some list $\{ \seq[m]{i} \}$, where in $\FSpi$ the arity of
$f$ is $m$. 
Every argument filtering $\pi$ induces a mapping from
$\TERMS$ to $\TERMSpi$, also denoted by $\pi$:
\begin{equation*}
 \pi(t) = \begin{cases}
t & \text{if $t$ is a variable} \\
\pi(t_i) & \text{if $t = f(\seq{t})$ and $\pi(f) = i$} \\
f(\pi(t_{k_1}),\dots,\pi(t_{k_m})) &
\text{if $t = f(\seq{t})$ and $\pi(f) = \{ \seq[m]{i} \}$} \tpkt
\end{cases} 
\end{equation*}

\begin{definition}
  Let $\pi$ denote an argument filtering, and $\gpop$ a polynomial
  path order. We define $s \gpopp t$ if and only if $\pi(s) \gpop
  \pi(t)$, and likewise $s \geqpopp t$ if and only if $\pi(s) \geqpop \pi(t)$.
\end{definition}

\begin{example}[Example \ref{ex:bits:widp} continued]
 \label{ex:bits:orient}
Let $\pi$ be defined as follows: $\pi(\halfs) = 1$ and $\pi(f) = \{1,\dots,n\}$ for each $n$-ary
function symbol other than $\halfs$. Compatibility of $\WIDP(\RSbits)$ with $\gpopp$ 
amounts to the following set of order constraints:
 \begin{alignat*}{6}
   && \mrk{\halfs}(0)& \gpop \mathsf{c_1} & \hspace{3ex} 
   &&\mrk{\bits}(0) & \gpop \mathsf{c_3} & \hspace{3ex} 
   &&\mrk{\halfs}(\ms(\ms(x)))& \gpop \mrk{\halfs}(x)
   \\
   && \mrk{\halfs}(\ms(0)) & \gpop \mathsf{c_2} & 
   && \mrk{\bits}(\ms(0)) & \gpop \mathsf{c_4} &
   && \mrk{\bits}(\ms(\ms(x)))& \gpop \mrk{\bits}(\ms(x))
 \end{alignat*}
In order to define a \POPSTAR\ instance $\gpop$, we set 
$\safe(\mrk{\bits})= \safe(\halfs) = \safe(\mrk{\halfs}) = \varnothing$ and $\safe(\ms) = \set{1}$.
Furthermore, we define an (admissible) precedence:
 $0 \ep \mathsf{c_1} \ep \mathsf{c_2} \ep \mathsf{c_3} \ep \mathsf{c_4}$.
The easy verification of $\WIDP(\RSbits) \subseteq {\gpopp}$ is left
to the reader.
\end{example}

\section{Dependency Pairs and Polynomial Path Orders} \label{s:dppop}

Motivated by Example~\ref{ex:bits:orient}, we show in this
section that the pair ($\geqpopp,\gpopp$) can play the role of a
\emph{safe} reduction pair, cf.~\cite{HirokawaMoser:2008,HirokawaMoser2:2008}.
Let $\RS$ be a TRS over a signature $\FS$ that is innermost terminating. In the sequel $\RS$ 
is kept fixed. Moreover, we fix some safe mapping $\safe$, an admissible precedence $\qp$, and an
argument filtering $\pi$. We refer to the induced $\POPSTAR$
instance by $\gpopp$.

We adapt $\safe$ to $\FSpi$ in
the obvious way: for each $\ftr{f} \in \FSpi$ with corresponding $f \in
\FS$, we define $\safe(\ftr{f}) \defi \safe(f) \cap \pi(f)$, 
and likewise $\normal(\ftr{f}) \defi \normal(f) \cap \pi(f)$.
Set $\Valpi \defi \TA(\CSpi,\VS)$. Based on $\FSpi$ we define the \emph{normalised
signature} $\FSnpi \defi \set{\fsn \mid f \in \FSpi}$ where the
arity of $\fsn$ is $\card{\normal(f)}$.
We extend $\qp$ to $\FSnpi$ by $\fsn \qp \gsn$ if and only if
$f \qp g$. Let $\ms$ be a fresh constant that is minimal with respect to $\qp$. 
We introduce the \emph{Buchholz norm} of $t$ (denoted 
as $\bN{t}$) a term complexity measure that fits well with the
definition of \POPSTAR. 
Set $\bN{t} \defi 1+ \max\set{n,\bN{t_1},\dots,\bN{t_n}}$
for $t=f(\seq{t})$ and $\bN{t} \defi 1$, otherwise.
In the following we define an embedding from the relative
rewriting relation $\irrewt{\RS}{\RSS}$ into $\gpopv_k$, such that $k$ depends
only on TRSs $\RS$ and $\RSS$. This embedding provides the technical
tool to measure the number of root steps in a given derivation through
the number of descent in $\gpopv_k$. Hence Proposition \ref{p:popseq}
becomes applicable to establishing our main result. This intuition
is cast into the next definition.
\begin{definition}\label{d:pred:int}
A \emph{predicative interpretation} is a pair of mappings
$(\ints,\intn)$ from terms to sequences $\SE(\FSnpi \cup
\set{\ms},\VS)$ defined as follows. We assume for $\pi(t) = f(\pi(t_1),\dots,\pi(t_n))$ that
$\safe(f) = \set{i_1,\dots,i_p}$ and $\normal(f) = \set{j_1,\dots,j_q}$.
  \begin{align*}
    \ints(t) & \defi
    \begin{cases}
       \emptys & \text{if $\pi(t) \in \Valpi$}, \\
       [\fsn(\intn(t_{j_1}),\dots,\intn(t_{j_q}))~\ints(t_{i_1})~\cdots~\ints(t_{i_p})] 
      & \text{if $\pi(t) \not\in \Valpi$.} 
    \end{cases}\\
    \intn(t) & \defi \ints(t) \cons \BN(t)
  \end{align*}
Here the function $\BN$ maps a term $t$ to the sequence $[\ms\cdots\ms]$ with
$\bN{\pi(t)}$ occurrences of the constant $\ms$.
\end{definition}
Note that as a direct consequence of the definitions we obtain
$\width(\intn(t)) = \bN{\pi(t)} + 1$ for all terms $t$.

\begin{lemma}\label{lem:slow:intn}
  There exists a polynomial $p$ such that 
  $\Slow(\intn(t)) \leqslant p(\size{t})$ for every basic
  term $t$. The polynomial $p$ depends only on $k$.
\end{lemma}
\begin{proof}
  Suppose $t=f(\seq{v})$ is a basic term with 
  $\safe(f) = \set{\seq[p]{i}}$ and $\normal(f) = \set{\seq[q]{j}}$.
  The only non-trivial
  case is when $\pi(t) \not \in \Valpi$. 
  Then
  $$
  \intn(t) = [u~\ints(v_{i_1}) \cdots \ints(v_{i_p})] \cons \BN(t)
  $$
  where $u = \fsn(\intn(v_{j_1}),\dots,\intn(v_{j_q}))$. Note that
  $\ints(v_i) = \emptys$ for $i \in \set{\seq[q]{i}}$.
  Let $\Ipop$ denote a polynomial interpretation fulfilling
  Proposition~\ref{p:popseq}.
  Using the assumption $\circ_\Ipop(\seq{m}) = \sum_{i=1}^n m_i + n$, it
  is easy to see that $\Slow(\intn(t))$ is bounded linear in $\bN{\pi(t)} \leqslant \size{t}$ 
  and $\interpret{\alpha}{u}$.
  As $\intn(v_j) = [[]~\ms\cdots\ms]$ 
  with $\bN{\pi(v_j)} \leqslant \size{t}$ occurrences
  of $\ms$, 
  $\Slow(\intn(v_j))$ is linear in $\size{t}$.
  Hence from Proposition~\ref{p:popseq} we conclude that
  $\Slow(\intn(t))$ is polynomially bounded in $\size{t}$.
\end{proof}

The next sequence of lemmas shows that the relative
rewriting relation $\irrewt{\RS}{\RSS}$ is embeddable into $\gpopv_k$.
\begin{lemma}\label{l:embed:val}
  Suppose $s \gpopp t$ such that $\pi(s\sigma)\in\Valpi$. Then
  $\ints(s\sigma) = \emptys = \ints(t\sigma)$ and
  $\intn(s\sigma) \gpopv_1 \intn(t\sigma)$.
\end{lemma}
\begin{proof}
  Let $\pi(s\sigma)\in\Valpi$, and 
  suppose $s \gpopp t$, i.e., $\pi(s) \gpop \pi(t)$ holds.
    Observe that since $\pi(s) \in \Valpi$ and due to our assumptions
    on safe mappings, only clause $\cl{1}$
    from the definition of $\gpop$ (or respectively $\gsq$) is applicable.
    And thus $\pi(t)$ is a subterm of $\pi(s)$ modulo the equivalence $\eqi$.
    We conclude $\pi(t\sigma)\in\Valpi$, and hence
    $\ints(s\sigma) = \emptys = \ints(t\sigma)$.
    Finally, notice that $\bN{\pi(s\sigma)} > \bN{\pi(t\sigma)}$ as
    $\pi(t\sigma)$ is a subterm of $\pi(s\sigma)$. Thus 
    $\intn(s\sigma) \gpopv_1 \intn(t\sigma)$ follows as well.
\end{proof}

To improve the clarity of the exposition, we concentrate on the curcial 
cases in the proofs of the following lemma. The interested reader is kindly referred
to~\cite{Avanzini:2009} for the full proof.
\begin{lemma} \label{l:embed:claim}
  Suppose $s \gsqp t$ such that $\pi(s\sigma) =
  f(\pi(s_{1}\sigma), \dots, \pi(s_{n}\sigma))$ with
  $\pi(s_{i}\sigma) \in \Valpi$ for $i \in \set{1,\dots,n}$. Moreover
  suppose $\normal(f) = \set{j_1,\dots,j_q}$.
  Then 
  $$\fsn(\intn(s_{j_1}\sigma), \dots,\intn(s_{j_q}\sigma)) \gppv_{3\cdot \bN{\pi(t)}} \intn(t\sigma)$$ 
  holds.
\end{lemma}
\begin{proof}
Note that the assumption implies that the argument filtering $\pi$ does not collapse $f$. 
We show the lemma by induction on $\gsqp$. We consider the subcase that $s \gsqp t$ follows  
as $t = g(\seq[m]{t})$, $\pi$ does not collapse on $g$, $f \succ g$, 
and $s \gsqp t_j$ for all $j \in \pi(g)$, cf.~Definition~\ref{d:pop2}(\ref{d:pop2:ii}).
We set $u \defi \fsn(\intn(s_{j_1}\sigma), \dots,\intn(s_{j_q}\sigma))$ and 
$k \defi 3\cdot \bN{\pi(t)}$ and first prove $u \gppv_{k - 1} \ints(t\sigma)$. 

If $\pi(t\sigma) \in \Valpi$, then $\ints(t\sigma) = \emptys$ is
minimal with respect to $\gppv_{k-1}$. Thus we are done.
Hence suppose $\normal(g) = \set{j'_1,\dots,j'_q}$, $\safe(g) = \set{i'_1,\dots,i'_p}$
and let
\begin{equation*}
\ints(t\sigma) = [\gsn(\intn(t_{j'_1}\sigma),\dots,
\intn(t_{j'_q}\sigma))~\ints(t_{i'_1}\sigma) \cdots \ints(t_{i'_p}\sigma)] \tpkt
\end{equation*}
We set $v \defi \gsn(\intn(t_{j'_1}\sigma),\dots,\intn(t_{j'_q}\sigma))$.
It suffices to show $u \gppv_{k-2} v$
and $u \gppv_{k-2} \ints(t_{j}\sigma)$ for $j \in \safe(g)$. 
Both assertions follow from the induction hypothesis.

Now consider $\intn(t\sigma) = [\ints(t\sigma)~\ms\cdots\ms]$
with $\bN{\pi(t\sigma)}$ occurrences of the constant $\ms$. 
Recall that $\width(\intn(t\sigma)) = \bN{\pi(t\sigma)} + 1$. 
Observe that $\fsn \succ \ms$. Hence to prove $u \gppv_{k} \ints(t\sigma)$ it suffices
to observe that $\width(u) + k > \bN{\pi(t\sigma)} + 1$ holds. For that 
note that $\bN{\pi(t\sigma)}$ is either $\bN{\pi(t_j\sigma)} + 1$ for some
$j \in \pi(g)$ or less than $k$. In the latter case, we are done. 
Otherwise $\bN{\pi(t\sigma)} = \bN{\pi(t_j\sigma)} + 1$. Then from 
the definition of $\gppv_k$ and the induction hypothesis 
$u \gppv_{3 \cdot \bN{\pi(t_j)}} \intn(t_j\sigma)$
we can conclude 
$\width(u) + 3 \cdot \bN{\pi(t_j)} > \width(\intn(t_j\sigma)) = \bN{\pi(t_j\sigma)} + 1$.
Since $k \geqslant 3 \cdot (\bN{\pi(t_j)} + 1)$, $\width(u) + k > \bN{\pi(t\sigma)} + 1$ follows.
\end{proof}

\begin{lemma}\label{l:embed:hlp}
  Suppose $s \gpopp t$ such that $\pi(s\sigma) =
  f(\pi(s_{1}\sigma), \dots, \pi(s_{n}\sigma))$ with
  $\pi(s_{i}\sigma) \in \Valpi$ for $i \in \set{1,\dots,n}$. Then
  for $\normal(f) = \set{j_1,\dots,j_q}$, 
  \begin{enumerate}
  \item \label{en:embed:hlp:a}
    $\fsn(\intn(s_{j_1}\sigma), \dots, \intn(s_{j_q}\sigma)) 
    \gpopv_{3\cdot \bN{\pi(t)}} \ints(t\sigma)$, and 
  \item \label{en:embed:hlp:b}
    $\fsn(\intn(s_{j_1}\sigma), \dots, \intn(s_{j_q}\sigma)) \cons \BN(s\sigma) 
    \gpopv_{3\cdot \bN{\pi(t)}} \intn(t\sigma)$.
  \end{enumerate}
\end{lemma}
\begin{proof}
The lemma is shown by induction on the definition of $\gpopp$.
For the following, we set $u = \fsn(\intn(s_{j_1}\sigma), \dots,\intn(s_{j_q}\sigma))$. 
Suppose $s \gpopp t$ follows due to Definition~\ref{d:pop3}(\ref{d:pop3:ii}). 
We set $k \defi 3\cdot \bN{\pi(t)}$. Let $\normal(g) = \set{j'_1,\dots,j'_q}$ and let
$\safe(g) = \set{i'_1,\dots,i'_p}$.

Property $(\ref{en:embed:hlp:a})$ is immediate for $\pi(t\sigma) \in \Valpi$, so 
assume otherwise. We see that $s \gsqp t_j$ for all $j \in \normal(g)$
and obtain $u \gppv_{k-1} \gsn(\intn(t_{j'_1}\sigma),\dots,\intn(t_{j'_q}\sigma))$ 
as in Lemma~\ref{l:embed:claim}.
Furthermore, $s \gpopp t_{j_0}$ for some $j_0 \in \safe(g)$
and by induction hypothesis: $u \gpopv_{k-1} \ints(t_{j_0}\sigma)$. 
To conclude property $(\ref{en:embed:hlp:a})$, it remains to verify $u \gppv_{k-1} \ints(t_j\sigma)$ 
for the remaining $j \in \safe(g)$.
We either have $s \gsqp t_j$ or $\pi(s_i) \superterm \pi(t_j)$ (for some $i$). 
In the former subcase we proceed as in the claim, and for the latter
we observe $\pi(t_j\sigma) \in \Valpi$, and thus 
$\ints(t_j\sigma) = \emptys$ follows.
This establishes property $(\ref{en:embed:hlp:a})$.

To conclude property $(\ref{en:embed:hlp:b})$, it suffices to show
$\width(u \cons \BN(s\sigma)) + k > \width(\intn(t\sigma))$, or
equivalently $\bN{\pi(s\sigma)} + 1 + k > \bN{\pi(t\sigma)}$.
The latter can be shown, if we proceed similar as in the claim. 
\end{proof}

Recall the definition of $\qrew{\QS}$ from Section~\ref{s:seq} and 
define $\QS \defi \{f(\seq{x}) \to \bot \mid f \in \DS\}$, and 
set $\vrew \defi {\qrew[\RS]{\QS}}$. 
As the normal forms of $\QS$ coincide with
$\Val$, $\vrew$ is the restriction of $\irew$, where arguments need to 
be values instead of normal forms of $\RS$.
From  Lemma \ref{l:embed:val} and \ref{l:embed:hlp} we 
derive an embedding of root steps $\vrewt[\RS]$.

Suppose the step $s \vrew[\RS] t$ takes place below the root.
Observe that $\pi(s) \not = \pi(t)$ need not hold in general. Thus 
we cannot hope to prove $\intn(s) \gpopv_k \intn(t)$. However,
we have the following stronger result.
\begin{lemma}\label{l:embed}
  There exists a uniform $k \in \NAT$ (depending only on $\RS$)
  such that if $\RS \subseteq {\gpopp}$ holds 
  then ${s \vrewt[\RS] t}$ implies ${\intn(s) \gpopv_k \intn(t)}$.
  Moreover, if $\RS \subseteq {\geqpopp}$ holds then 
   ${s \vrew[\RS] t}$ implies ${\intn(s) \geqpopv_k \intn(t)}$. 
\end{lemma}
\begin{proof}
  We consider the first half of the assertion.
  Suppose $\RS \subseteq {\gpopp}$ and $s \vrewt[\RS] t$, that is for some rule 
  ${f(\seq{l}) \to r} \in \RS$ and 
  substitution $\ofdom{\sigma}{\VS \to \Val}$ we have 
  $s = f(l_1\sigma,\dots,l_n\sigma)$ and  $t = r\sigma$. 
  Depending on whether $\pi$ collapses $f$, the property either
  directly follows from Lemma \ref{l:embed:val} or is a 
  consequence of Lemma $\ref{l:embed:hlp}(\ref{en:embed:hlp:b})$. 
  
  In order to conclude the second half of the assertion, 
  one performs induction on the rewrite context. In addition, 
  one shows that for the special case $\ints(s) \eqi \ints(t)$, 
  still $\bN{\pi(s)} \geqslant \bN{\pi(t)}$ holds.
  From this the lemma follows.
\end{proof}

For constructor TRSs, we can simulate $\irew[\RS]$
using $\vrew[\RS]$. We extend $\RS$ with suitable rules $\V(\RS)$,
which replace normal forms that are not values 
by some constructor symbol. To simplfy the argument we re-use the
symbol $\bot$ from above.
We define the TRS $\V(\RS)$ as
\begin{equation*}
  \V(\RS) \defi \set{ f(\seq{t}) \to \bot \mid f(\seq{t}) \in
    {\NF(\RS) \cap \TA(\FS)} \text{ and } f \in \DS}\tpkt
\end{equation*}
Moreover, we define $\nfV[\RS]{t} \defi \nf[\V(\RS)]{t}$.
Observe that $\nfV[\RS]{\cdot}$ is well-defined since $\V(\RS)$ is 
confluent and terminating. 

\begin{lemma}\label{l:v:simulation:relative}
  Let $\RS \cup \RSS$ be a constructor TRS. 
  Define $\RSS' \defi \RSS \cup \V(\RS \cup \RSS)$.
  For $s \in \TA(\FS)$, 
  \begin{equation*}
    {s \irrewt{\RS}{\RSS} t} \quad \text{implies} \quad {\nfV[\RS \cup \RSS]{s}
    \vrrewt{\RS}{\RSS'} \nfV[\RS \cup \RSS]{t}} \tkom 
  \end{equation*}
  where ${\vrrew{\RS}{\RSS'}}$ abbreviates 
  ${\vrss[\RSS'] \cdot \vrew[\RS] \cdot \vrss[\RSS']}$.
\end{lemma}
\begin{proof}
  It is easy to see that
  ${s \irew[\RS] t}$ implies 
  $\nfV[\RS]{s} \vrew[\RS] \cdot \vrsn[\V(\RS)] \nfV[\RS]{t}$.
  Suppose $s \irrewt{\RS}{\RSS} t$, then 
  there exist ground terms $u$ and $v$ such that 
  $s \irss[\RSS] u \irewt[\RS] v \irss[\RSS] t$. 
  Let $\nfV{t} \defi \nfV[\RS \cup \RSS]{t}$. 
  From the above,
  $\nfV{s} 
  \vrss[\RSS']
  \nfV{u} 
  \vrewt[\RS] \cdot \vrss[\RSS']
  \nfV{v} 
  \vrss[\RSS']
  \nfV{t}$ follows as desired. 
\end{proof}

Suppose $\RS \subseteq {\gpopp}$ and $\RSS \subseteq {\geqpopp}$ holds.
Together with Lemma \ref{l:embed}, the above simulation establishes 
the promised embedding of $\irrewt{\RS}{\RSS}$ into $\gpopv_k$.

\begin{lemma}\label{l:relstep:root}
  Let $\RS \cup \RSS$ be a constructor TRS, and suppose 
  $\RS \subseteq {\gpopp}$ and ${\RSS} \subseteq {\geqpopp}$ hold.
  Then for $k$ depending only on $\RS$ and $\RSS$ and $s \in \TA(\FS)$,
  we have 
  \begin{equation*}
    {s \irrewt{\RS}{\RSS} t} \quad \text{implies} \quad {\intn(\nfV{s}) \gpopv_k^+ \intn(\nfV{t})} \tpkt
  \end{equation*}
\end{lemma}
\begin{proof}
  Consider a step $s \irrewt{\RS}{\RSS} t$ and set $\nfV{t} \defi \nfV[\RS \cup \RSS]{t}$.
  By Lemma \ref{l:v:simulation:relative}
  there exist terms $u$ and $v$ such that
  $\nfV{s} \vrss[\RSS \cup \V(\RS \cup \RSS)] u \vrewt[\RS] v \vrss[\RSS \cup \V(\RS \cup \RSS)] \nfV{t}$.
  %
  Since $\RS \subseteq {\gpopp}$ holds, by Lemma \ref{l:embed}
  $\intn(u)  \gpopv_{k_1} \intn(v)$ follows. 
  Moreover from $\RSS \subseteq {\geqpopp}$ together with Lemma \ref{l:embed} we conclude that
  $r_1 \vrew[\RSS \cup \V(\RS \cup \RSS)] r_2$ implies $\intn(r_1) \geqpopv_{k_2} \intn(r_2)$. Here it 
  suffices to see that steps from  $\VS(\RS \cup \RSS)$ are easy to
  embed
  into $\geqpopv_{k_2}$ using the predicative interpretation $\intn$ independent of $k_2$.
  %
  In both cases $k_1$ and $k_2$ depend only on $\RS$ and $\RSS$ respectively; set
  $k \defi \max\set{k_1,k_2}$. 
  In sum we have $\intn(\nfV{s}) \geqpopv_k^* \intn(u) \gpopv_k \intn(v) \geqpopv_k^* \intn(\nfV{t})$,
  employing ${\gpopv_{l_1}} \subseteq {\gpopv_{l_2}}$ for $l_1 \leqslant l_2$.
  It is an easy exercise to show that ${{\gpopv_k} \cdot {\eqi}} \subseteq {\gpopv_k}$
  and likewise ${{\eqi} \cdot {\gpopv_k}} \subseteq {\gpopv_k}$ holds.
  Hence the lemma follows.
\end{proof}

\begin{theorem}\label{t:relstep:root}
  Let $\RS \cup \RSS$ be a constructor TRS, and suppose 
  $\RS \subseteq {\gpopp}$ and ${\RSS} \subseteq {\geqpopp}$ holds.
  Then there exists a polynomial $p$ depending only on $\RS \cup \RSS$
  such that for any basic and ground term $t$,
  $\dl(t, \irrewt{\RS}{\RSS}) \leqslant p(\size{t})$.
\end{theorem}
\begin{proof}
Assume $t \not \in \NF(\RS \cup \RSS)$, otherwise $\dl(t,
\irrewt{\RS}{\RSS})$ is trivially bounded. 
Moreover $t$ is a basic term, hence $\nfV[\RS \cup \RSS]{t} = t$.
  From Lemma~\ref{l:relstep:root} we infer that 
  $\dl(t, \irrewt{\RS}{\RSS}) \leqslant \Slow(\intn(\nfV[\RS \cup
  \RSS]{t})) = \Slow(\intn({t}))$ for some $k$, where the latter 
  is polynomially bounded in $\size{t}$ and the polynomial only
  depends on $k$, cf. Lemma \ref{lem:slow:intn}. Finally $k$ depends only on $\RS \cup \RSS$.
\end{proof}

Suppose $\RS$ is a constructor TRS, and let $\PS$ denote the set of weak innermost 
dependency pairs. 
For the moment, suppose that all compound symbols of $\PS$ are nullary.
Provided that $\PS$ is non-duplicating and compatible with some SLI, 
as a consequence of the above theorem paired with Proposition \ref{p:dp},
the inclusions $\PS \subseteq {\gpopp}$ and $\URs(\PS) \subseteq {\geqpopp}$ certify that
$\rcRi$ is polynomially bounded. 
Observe that for the application of $\gpopp$ and $\geqpopp$ in the context of
$\PS$ and $\URs(\PS)$, we alter Definitions~\ref{d:pop2} and~\ref{d:pop3} such 
that $f \in \mrk{\DS}$ is demanded.
\begin{example}
  [Example \ref{ex:bits:orient} continued]
  Reconsider the TRS $\RSbits$, and let
  $\PS$ denote $\WIDP(\RSbits)$ as drawn in Example \ref{ex:bits:widp}.
  By taking the SLI $\AS$ with
  $0_\AS = 0$, $\ms_\AS(x) = x + 1$ and $\halfs_\AS(x) = x  + 1$
  we obtain $\URs(\PS) \subseteq {\gord{\AS}}$ and moreover, observe that
  $\PS$ is both non-duplicating and contains only nullary compound symbols.
  In Example \ref{ex:bits:orient} we have seen that $\PS \subseteq
  {\gpopp}$ holds.  Similar, $\URs(\WIDP(\RSbits)) \subseteq
  {\geqpopp}$ can easily be shown.
  From the above observation we thus conclude a polynomial
  runtime-complexity of $\RSbits$.
\end{example}

The assumption that all compound symbols from $\PS$ need
to be nullary is straightforward to lift, but technical. Hence,
we do not provide a complete proof here, but only indicate the
necessary changes. The formal construction can be found in the Appendix.

Note that in the general case, it does not suffice to embed root steps of $\PS$ into 
$\gpopv_k$, rather we have to embed steps of form
$C[\mrk{s_1},\dots,\mrk{s_i},\dots,\mrk{s_n}] \vrew[\PS] 
C[\mrk{s_1},\dots,\mrk{t_i},\dots,\mrk{s_n}]$ with $C$ being a context
built from compound symbols. 
As first measure we require that the argument filtering $\pi$ is \emph{safe}
\cite{HirokawaMoser:2008}, that is $\pi(c) = [1,\dots,n]$ for each compound symbol
$\m{c}$ of arity $n$.
Secondly, we adapt the predicative interpretation $\intn$ in such a
way that compound symbols are interpreted as sequences, 
and their arguments by the interpretation $\intn$.
This way, a proper embedding using $\intn$ requires 
$\intn(\mrk{s_i}) \gpopv_k \intn(\mrk{t_i})$ 
instead of $\ints(\mrk{s_i}) \gpopv_k \ints(\mrk{t_i})$.

\begin{theorem}\label{t:widp}
    Let $\RS$ be a constructor TRS, and let $\PS$ denote the set of weak innermost 
    dependency pairs. 
    Assume $\PS$ is non-duplicating, and suppose ${\URs(\PS)} \subseteq {>_\AS}$ 
    for some SLI $\AS$. Let $\pi$ be a safe 
    argument filtering.
    If $\PS \subseteq {\gpopp}$ and $\URs(\PS) \subseteq {\geqpopp}$ then
    $\rcRi$ is polynomially bounded. 
\end{theorem}

Above it is essential that $\RS$ is a constructor TRS. This
even holds when $\POPSTAR$ is applied directly.
\begin{example}\label{ex:counterexample}
Consider the TRS $\RS_{\cexp}$ below:
\begin{equation*}
\begin{array}{c}
    \cexp(x) \to \me(\mg(x)) \qquad
     \me(\mg(\ms(x)))  \to \cdup_1(\mg(x)) \qquad 
     \mg(\mN)  \to \mN \BOT\\
     \cdup_1(x)  \to \cdup_2(\me(x),x) \qquad\qquad
     \cdup_2(x,y)  \to \cpr(x,\me(y))
   \end{array}  
\end{equation*}
The above rules
are oriented (directly) by $\gpop$ induced by $\safe$ and $\qp$ such that:
(i) the argument position of $\mg$ and $\cexp$ are normal, the
remaining argument positions are safe, and (ii)
$\cexp \succ \mg \succ \cdup_1 \succ \cdup_2  \succ \me \succ \cpr
\succ \mN$.
On the other hand, $\RS_{\cexp}$ admits at least exponential 
innermost runtime-complexity, as 
for instance $\cexp(s^n(\mN))$ normalizes
in exponentially (in $n$) many innermost rewrite steps.
\end{example}

To overcome this obstacle, we adapt the definition of $\gpop$
in the sense that we refine the notion of defined function symbols
as follows. Let $\GC$ denote the least set containing $\CS$
and all symbols appearing in arguments to left-hand sides
in $\RS$. Moreover, set $\GD \defi \FS \setminus \GC$ and
set $\Val \defi \TA(\GC,\VS)$.
Then in order to extend Theorem~\ref{t:widp} to non-constructor TRS
it suffices to replace $\DS$ by $\GD$ and $\CS$ by $\GC$ in all 
above given definitions and arguments (see~\cite{Avanzini:2009} for the
formal construction).
Thus the next theorem follows easily from combining Proposition~\ref{p:dg}
and Theorem~\ref{t:widp}. Note that this theorem can be easily extended
so that in each path different termination techniques (inducing polynomial
runtime complexity) are employed, see~\cite{HirokawaMoser2:2008} and Section~\ref{s:exps}. 
\begin{theorem} \label{t:widg}
  Let $\RS$ be a TRS.
  Let $\GS$ denote the weak innermost dependency graph, 
  and let $\FS = \GD \uplus \GC$ be separated as above.
  Suppose for every path $(\PS_1,\ldots,\PS_n)$  in $\PG{\GS}$ there exists
  an SLI $\AS$ and a pair 
  $(\geqpopp,\gpopp)$ based on a safe argument filtering $\pi$
  such that
  (i) $\URs(\PS_1 \cup \cdots \cup \PS_n) \subseteq {>_{\AS}}$
  (ii) $\PS_1 \cup \cdots \cup \PS_{n-1} \cup \URs(\PS_1 \cup \cdots \cup \PS_n)
  \subseteq {\geqpopp}$, and
  (iii) $\PS_n \subseteq {\gpopp}$ holds.
  Then $\rcRi$ is polynomially bounded. 
\end{theorem}

The next theorem establishes that \POPSTAR\ in conjunction with 
(weak innermost) dependency pairs induces polytime computability of the function
described through the analysed TRS. We kindly refer the reader to the Appendix
for the proof.
\begin{theorem}
  \label{t:polytime}
  Let $\RS$ be an orthogonal, $S$-sorted 
  and completely defined constructor TRS such that the underlying 
  signature is simple.
  Let $\PS$ denote the set of weak innermost dependency pairs. 
  Assume $\PS$ is non-duplicating, and suppose ${\URs(\PS)} \subseteq {>_\AS}$ 
  for some SLI $\AS$.
  If $\PS \subseteq {\gpopp}$ and $\URs(\PS) \subseteq {\geqpopp}$
  then
  the functions computed by $\RS$ are computable in polynomial time.
\end{theorem}
Here \emph{simple} signature \cite{Marion:2003} 
essentially means that the size
of any constructor term depends polynomially on its depth. 
Such a restriction is always necessary in this context. A detailed
account is given in the Appendix (see alo~\cite{Marion:2003}).
This restriction is also responsible for the introduction of sorts.

\section{Experimental Results} \label{s:exps}

\newcommand{\MTHD}[1]{\textsf{#1}}
\newcommand{\PP}{\MTHD{PP}}
\renewcommand{\P}{\MTHD{P}}
\newcommand{\MP}{\MTHD{MP}}
\newcommand{\M}{\MTHD{M}}
\newcommand{\WIDGC}{\MTHD{WIDG}}
\newcommand{\WIDPC}{\MTHD{WIDP}}
\newcommand{\DIRECT}{\MTHD{DIRECT}}

All described techniques have been incorporated into
the \emph{Tyrolean Complexity Tool} $\TCT$, 
an open source complexity analyser%
\footnote{Available at \url{http://cl-informatik.uibk.ac.at/software/tct}.}.
We performed tests on two testbeds: \textbf{T} constitutes of the 
$1394$ examples from the Termination Problem Database Version 5.0.2
that were used in the runtime-complexity category of the termination competition 2008\footnote{%
See~\url{http://termcomp.uibk.ac.at}.}.
Moreover, testbed \textbf{C} is the restriction of testbed \textbf{T} 
to constructor TRSs ($638$ in total). 
All experiments were conducted on a machine that is identical 
to the official competition server
($8$ AMD Opteron${}^\text{\textregistered}$ 885 dual-core processors
with 2.8GHz, $8\text{x}8$ GB memory). As timeout we use 5 seconds.
We orient TRSs using 
$\gpopp$ by encoding the constraints on precedence and so forth 
in \emph{propositional logic} (cf. \cite{Avanzini:2009} for details), employing
$\mathsf{MiniSat}$~\cite{EenSorensson:2003}
for finding satisfying assignments.
In a similar spirit, we check compatibility with SLIs via translations
to SAT. In order to derive an estimated dependency graph, we use the function $\mathsf{ICAP}$
(cf.~\cite{GieslThiemannSK:2005}).

Experimental findings are summarised in Table~\ref{tbl:exp1}.%
\footnote{See~\url{http://cl-informatik.uibk.ac.at/~zini/rta09} for
  extended results.}
In each column, we highlight the total on yes-, maybe- and
timeout-instances. Furthermore, we annotate average 
times in seconds. In the first three columns we contrast \POPSTAR\ as direct technique 
to \POPSTAR\ as base to (weak innermost) dependency pairs. I.e., the columns $\WIDPC$ and $\WIDGC$ show results
concerning Proposition \ref{p:usable} together with Theorem~\ref{t:widp} or Theorem~\ref{t:widg} respectively.

In the remaining four columns we assess the power of Proposition~\ref{p:usable} and~\ref{p:dg}
in conjunction with different base orders, thus verifying that the use of \POPSTAR\ in this context
is independent to existing techniques.
Column \P~asserts that the different paths are handled by 
\emph{linear and quadratic restricted interpretations}~\cite{HirokawaMoser:2008}.
In column~\PP, in addition \POPSTAR~is employed.
Similar, in column \M~\emph{restricted matrix interpretations} 
(that is matrix interpretations~\cite{EndrullisWaldmannZantema:2008}, where 
constructors are interpreted by triangular matrices)
are used to handle different paths. 
Again column \MP~extends column \M~with \POPSTAR.
Note that all methods induce polynomial innermost runtime complexity.
\newcommand{\tm}[1]{\bf{\tiny{/#1}}}
\renewcommand{\c}[1]{{\small{#1}}}
\newcommand{\spc}{@{\hspace{1.8mm}}}
\begin{table}[h]
  \centering
  \begin{tabular}{c@{\hspace{1ex}}|@{\hspace{1pt}}|l@{\hspace{2mm}}
      r@{}l \spc r@{}l \spc r@{}l \spc\spc r@{}l \spc r@{}l \spc r@{}l \spc r@{}l}
    \multicolumn{2}{c}{\TOP }            & \multicolumn{6}{c}{polynomial path orders} 
                                         & \multicolumn{8}{c}{dependency graphs mixed} \\
    \multicolumn{2}{c}{\BOT }            & \multicolumn{2}{c}{\DIRECT} & \multicolumn{2}{c}{\WIDPC} & \multicolumn{2}{c}{\WIDGC}
                                         & \multicolumn{2}{c}{\P}   & \multicolumn{2}{c}{\PP}      & \multicolumn{2}{c}{\M} & \multicolumn{2}{c}{\MP}  \\
    \hline
    \textbf{T} 
     & Yes \TOP                              & \c{46}   & \tm{0.03}     & \c{69}   & \tm{0.09}    & \c{80}    & \tm{0.07} 
                                             & \c{198}  & \tm{0.54}     & \c{198}  & \tm{0.51}    & \c{200}   & \tm{0.63} & \c{207}   & \tm{0.48} \\
     & Maybe                                 & \c{1348} & \tm{0.04}     & \c{1322} & \tm{0.10}    & \c{1302}  & \tm{0.14}
                                             & \c{167}  & \tm{0.77}     & \c{170}  & \tm{0.82}    & \c{142}   & \tm{0.61} & \c{142}    & \tm{0.63}    \\
     & Timeout \BOT                          & \c{0}    &               & \c{3}    &              & \c{12}    &
                                             & \c{1029} &               & \c{1026} &              & \c{1052}  &           & \c{1045}          \\
\hline
    \textbf{C} 
     & Yes \TOP                              & \c{40}   & \tm{0.03}     & \c{48}   & \tm{0.08}    & \c{55}    & \tm{0.05}    
                                             & \c{99}   & \tm{0.40}     & \c{100}  & \tm{0.38}    & \c{98}   & \tm{0.26} & \c{105}    & \tm{0.23}    \\
     & Maybe                                 & \c{598}  & \tm{0.05}     & \c{587}  & \tm{0.10}    & \c{576}   & \tm{0.13}
                                             & \c{143}  & \tm{0.72}     & \c{146}   & \tm{0.77}    & \c{119}    & \tm{0.51} & \c{119}    & \tm{0.54}    \\
     & Timeout \BOT                          & \c{0}    &               & \c{3}    &              & \c{7}     &
                                             & \c{396}  &               & \c{392}  &              & \c{421}   &           & \c{414} \\
    \hline
  \end{tabular}
\smallskip
\caption{Experimental Results}
\label{tbl:exp1}
 \end{table}

Table \ref{tbl:exp1} reflects 
that the integration of \POPSTAR~in the context of (weak) dependency pairs,
significantly extends the direct approach. Worthy of note, the extension of
\cite{AvanziniMoser:2008} with quasi-precedences alone gives 5
additional examples.
As advertised, \POPSTAR~is incredibly fast in all settings. 
Consequently, as evident from the table, polynomial path orders 
team well with existing techniques, without affecting overall
performance: notice that due to the additional of~\POPSTAR\ the
number of timeouts is reduced.

\section{Conclusion} 
\label{s:concl}

In this paper we study the runtime complexity of rewrite systems. We combine
two recently developed techniques in the context of complexity analysis: weak innermost
dependency pairs and polynomial path orders. 
If the conditions of our main result are met, we can conclude the
innermost polynomial runtime complexity of the studied term rewrite
system. And we obtain that the function defined are 
\emph{polytime computable}.
We have implemented the technique and experimental evidence clearly indicates the power
and in particular the efficiency of the new method. 

\appendix

\newtheorem*{theorempoly}{Theorem}{\bf}{\it}
\newtheorem*{theoremwidp}{Theorem}{\bf}{\it}

\section{Appendix}

Below we present the missing proofs of 
Theorem~\ref{t:widp} and Theorem~\ref{t:polytime} respectively.

As mentioned in Section~\ref{s:dppop}, we now 
introduce an \emph{extended predicative interpretation} 
whose purpose is to interpret
compound symbols as sequences, and their arguments via the 
interpretation $\intn$. 
\begin{definition}\label{d:intq}
  The \emph{extended predicative interpretation} 
  $\intq$ from terms $\TERMS$ to sequences $\SE(\FSnpi \cup
  \set{\ms},\VS)$ is defined as follows:
  if $t = \cs(\seq{t})$ and $\cs \in \Ccom$ then 
  $\intq(t) \defi [\intq(t_1)~\cdots~\intq(t_n)]$, and 
  otherwise $\intq(t) \defi [\intn(t)]$.
\end{definition}

Following \cite[Section 6.5]{TeReSe}, we briefly recall \emph{typed rewriting}.
Let $S$ be a finite set representing the set of \emph{types} or \emph{sorts}.
An \emph{$S$-sorted set $A$} is a family of sets $\set{A_s \mid s \in S}$ 
such that all sets $A_s$ are pairwise disjoint. 
In the following, we suppose that $\VS$ denotes an $S$-sorted set of variables.
An \emph{$S$-sorted signature $\FS$} is like a  signature, 
but the \emph{arity} of $f \in \FS$ is defined by $\ar(f) = (s_1,\dots,s_n)$ 
for $s_1,\dots,s_n \in S$.
Additionally, each symbol $f \in \FS$ is associated with a sort $s \in S$, 
called the \emph{type of $f$} and denoted by $\type(f)$.
We adopt the usual notion and write $\ofdom{f}{(s_1,\dots,s_n) \to s}$
when $\ar(f) = (s_1,\dots,s_n)$ and $\type(f) = s$.
The \emph{$S$-sorted set of terms $\TERMS_S$} consists of the sets
$\TERMS_s$ for $s \in S$, where $\TERMS_s$ is inductively 
defined by (i) $\VS_s \subseteq \TERMS_s$, and
(ii) $f(\seq{t}) \in \TERMS_s$ for all function symbols $f \in \FS$,
$\ofdom{f}{(s_1,\dots,s_n) \to s}$
and terms $t_i \in \TERMS_{s_i}$ for $i \in \set{1,\dots,n}$.
We say that a term $t$ is \emph{well-typed} if $t \in \TERMS_s$ for some sort $s$.
An $S$-sorted term rewrite system $\RS$ is a TRS such that for ${l \to r} \in \RS$, it holds that
$l,r \in \TERMS_s$ for some sort $s \in S$. As a consequence, for $s \in \TERMS_s$ and $s \rew t$, 
we have that $t \in \TERMS_s$. 

\begin{example}\label{ex:a:1}
  Let $S = \set{\Sbool,\Slist,\Snat,\Spair}$.
  The $S$-sorted rewrite system $\RSe$ is 
  given by the following rules:
  \begin{alignat*}{4}
    \fs(\ms(x)) & \to \cns(\prs(x,\gs(x)), \fs(x)) & \hspace{3ex} 
    \gs(\ms(x)) & \to \gs(x) \\
    \fs(0) & \to \cnil &
    \gs(0) & \to \tts
  \end{alignat*}
  Here we assign arities and sorts as follows:
  for the constructors we set
  $\ofdom{0}{\Snat}$, 
  $\ofdom{\ms}{\Snat \to \Snat}$, 
  $\ofdom{\prs}{(\Snat, \Sbool) \to \Spair}$,
  $\ofdom{\tts}{\Sbool}$,
  $\ofdom{\cnil}{\Slist}$,
  $\ofdom{\cns}{(\Spair,\Slist) \to \Slist}$;
  for the defined symbols we set
  $\ofdom{\fs}{\Snat \to \Slist}$ and
  $\ofdom{\gs}{\Snat \to \Sbool}$.
\end{example}

A \emph{simple} signature \cite{Marion:2003} is a sorted signature such that 
each sort has a finite \emph{rank} $r$ in the following sense:
the sort $s$ has rank $r$ if for every constructor 
$\ofdom{c}{(\seq{s}) \to s}$, the rank of each sort $s_i$ is less
than the rank of $s$, except for at most one sort which can 
be of rank $r$.
Simple signatures allow the definition of enumerated datatypes and inductive datatypes like words and 
lists but prohibit for instance the definition of tree structures. 
Observe that the signature underlying $\RSe$ from Example~\ref{ex:a:1}
is simple.
A crucial insight is that sizes of values formed from a simple signature 
can be estimated polynomially in their depth.
The easy proof of the following proposition can be found in \cite[Proposition~17]{Marion:2003}.
\begin{proposition}\label{prop:ss}
  Let $\CS$ be a set of constructors from a simple signature $\FS$. 
  There  exists a constant $d \in \NAT$ such that 
  for each term $t \in \TA(\CS,\VS)_S$ whose rank is $r$, 
  $\size{t} \leqslant d^r \cdot \depth(t)^{r+1}$.
\end{proposition}
In order to give a polytime algorithm for the functions computed 
by a TRS, it is essential that sizes of reducts do not exceed 
a polynomial bound with respect to the size of the start term.
Recall that approximations $\gpopv_k$ tightly control the size growth
of terms. For simple signatures, we can exploit this property for a
space-complexity analysis. 
Although predicative interpretations remove values, by the above
proposition sizes of those can be estimated based on the Buchholz-norm 
record in $\intn$. 
And so we derive the following Lemma, essential for the proof of Theorem~\ref{t:polytime}. 
\begin{lemma}\label{l:slow:sizebound}
  Let $\FS$ be a simple signature. 
  There exists a (monotone) polynomial $p$ depending only on $\FS$ such that 
  for each well-typed term $t \in \TA(\FS,\VS)_s$, 
  $\size{t} \leqslant p(\Slow(\intq[](t)))$.
\end{lemma}
\begin{proof}
The Lemma follows as:
(i) for all sequences $s \in \SE$, $\size{s} \leqslant \Slow(s) + 1$, and 
(ii) for all terms $t \in \TA(\FS,\VS)_s$, $ \size{t} \leqslant c \cdot \size{\intq[](t)}^d $
for some uniform constants $0 < c,d\in \NAT$.
These properties are simple to verify: property (i) follows from induction on $s$
where we employ for the inductive step that $f(\seq{s}) \gpopv_k [\sexpr{s}]$
and $\Slow([\sexpr{s}]) = \sum_{i=1}^n \Slow(s_i) + n$. 
For property (ii), 
set $d = r + 2$ where $r$ is 
the maximal rank of a symbol in $\CS$, and set $c = e^r$ where $e$ 
is as given from Proposition~\ref{prop:ss}.
First one shows by a straight forward 
induction on $t$ that 
$\size{t} \leqslant c \cdot (\size{\ints[](t)} \cdot \bN{t}^{r+1})$ 
(employing Proposition~\ref{prop:ss} and $\depth(t) \leqslant \bN{t}$).
As $\size{\ints[](t)} < \size{\intn[](t)}$ and
$\bN{t} < \size{\intn[](t)}$, 
we derive $\size{t} < c \cdot \size{\intn[](t)}^{d}$.
By induction on the definition of $\intq[]$ we finally obtain property
(ii). 
\end{proof}

Let $\RS$ be a (not necessarily $S$-sorted) TRS that is innermost terminating.
In the sequel, we keep $\RS$ fixed.
In order to exploit Lemma~\ref{l:slow:sizebound} for an analysis by means of 
weak innermost dependency pairs, we introduce the notion of 
\emph{type preserving weak innermost dependency pairs}.

\begin{definition}
  If $l \to r \in \RS$ and $r = \Ctx{\seq{u}}_{\DS}$ then 
  $\mrk{l} \to \mc(\mrk{u_1},\ldots,\mrk{u_n})$ 
  is called a \emph{type preserving weak innermost dependency pair} of
  $\RS$. 
  Here, the \emph{compound symbol} $\mc$ is supposed to be fresh.
  We set $\repr{\mc} \defi C$ and say that $\mc$ \emph{represents} the context $C$. 
  The set of all type preserving 
  weak innermost dependency pairs is denoted by $\WIDP(\RS)$.
\end{definition}
We collect all compound symbols appearing 
in $\TPWIDP(\RS)$ in the set $\Ccom$.

\begin{example}[Example~\ref{ex:a:1} continued]
  \label{ex:a:2}
  Reconsider the rewrite system $\RSe$ given in Example~\ref{ex:a:1}.
  The set $\TPWIDP(\RSe)$ is given by
  \begin{alignat*}{4}
    \mrk{\fs}(\ms(x)) & \to \cs_1(\mrk{\gs}(x), \mrk{\fs}(x)) & \hspace{3ex} 
    \mrk{\gs}(\ms(x)) & \to \cs_3(\mrk{\gs}(x)) \\
    \mrk{\fs}(0) & \to \cs_2 &
    \mrk{\gs}(0) & \to \cs_4
  \end{alignat*}
  The constant $\cs_3$ represents for instance the empty context, and
  the constant $\cs_1$ represents the context $\repr{\cs_1} = \cns(\prs(x,\hole), \hole)$. 
\end{example}

\begin{lemma}\label{l:tpwidp:preserve}
  Let $\RS$ be an $S$-sorted TRS such that the underlying signature
  $\FS$ is simple. 
  Then $\TPWIDP(\RS) \cup \URs(\WIDP(\RS))$ is an $S$-sorted TRS, 
  and the underlying signature $\mrk{\FS} \cup \Ccom$ a simple signature. 
\end{lemma}
\begin{proof}
  To conclude the claim, it suffices to type the marked and compound symbols appropriately.
  For each rule ${\mrk{f}(\seq{l}) \to \cs(\mrk{r_1},\dots,\mrk{r_n})}
  \in \TPWIDP(\RS)$ 
  we proceed as follows:
  we set $\ar(\mrk{f}) \defi \ar(f)$ and $\type(\mrk{f}) \defi \type(f)$. Moreover, 
  we set $\ar(\mc) \defi (\map[m]{\type}{r})$ and $\type(\mc) \defi \type(f)$.
  It is easy to see that since $\RS$ is $S$-sorted, $\TPWIDP(\RS) \cup
  \URs(\TPWIDP(\RS))$ is $S$-sorted
  too. 
\end{proof}
Note that the above lemma fails for weak innermost dependency
pairs: consider the rule $\fs(x) \to \ds(\gs(x))$, where $\fs$ and $\gs$ are
defined symbols and $\ds$ is a constructor. Moreover, suppose
$\ofdom{\fs}{\mathsf{s_2} \to \mathsf{s_1}}$, 
$\ofdom{\gs}{\mathsf{s_2} \to \mathsf{s_3}}$
and $\ofdom{\ds}{\mathsf{s_3} \to \mathsf{s_1}}$.
Then we cannot type the corresponding weak innermost dependency pair
$\mrk{\fs}(x) \to \mrk{\gs}(x)$ as above because (return-)types of $\mrk{\fs}$ and $\mrk{\gs}$
differ.

As for practical all termination techniques, compatibility 
of weak innermost dependency pairs with 
polynomial path orders also yield
compatibility of type preserving weak
innermost dependency pairs.
Moreover, from the definition we immediately see that 
$\dl(\mrk{t},\irrew{\TPWIDP(\RS)}{\US}) = \dl(\mrk{t},\irrew{\WIDP(\RS)}{\US})$
with $\US = \URs(\WIDP(\RS))$ and  basic term $t$.
And so it is clear that in order to proof Theorem~\ref{t:widp} and
Theorem~\ref{t:polytime}, 
$\WIDP(\RS)$ can safely be replaced by $\TPWIDP(\RS)$.
We continue with the proof of Theorem~\ref{t:widp}.

\subsection{Proof of Theorem~\ref{t:widp}}\label{s:proof:widp}

Let $\CCtx$ abbreviate the set of contexts 
$\TA(\Ccom \cup \set{\hole},\VS)$ build from compound symbols.
Set $\PS = \TPWIDP(\RS)$ and $\US = \URs(\WIDP(\RS))$.
In order to highlight the correspondence
between $\irew$ and $\irrew{\PS}{\US}$, 
we extend the notion of \emph{representatives}.
\begin{definition}\label{d:reprs}
  Let $C \in \CCtx$. We define $\reprs{C}$ as the least set of
  (ground) contexts such that
  (i) if $C = \hole$ then $\hole \in \reprs{C}$, and 
  (ii) if $C = \cs(\seq{C})$,
  $C'_i \in \reprs{C_i}$ and 
  $\sigma$ is a substitution from all variables in $\repr{\cs}$ to
  ground normal forms of $\RS$ then  
  $(\repr{\cs}\sigma)[C'_1,\dots,C'_n] \in \reprs{C}$.
\end{definition}

\begin{example}[Example~\ref{ex:a:2} continued]
  \label{ex:a:3}
  Reconsider the TRS $\RSe$ from Example~\ref{ex:a:1}, together
  with $\TPWIDP(\RSe)$ as given in Example~\ref{ex:a:2}.
  Consider the step 
  $\mf(\ms(0)) \rew[\RSe] \cns(\prs(0,\gs(0)), \fs(0))$ 
  and the corresponding dependency pair step
  \begin{equation*}
   \mrk{\mf}(\ms(0)) \rew[\TPWIDP(\RSe)] \cs_1(\mrk{\gs}(0), \mrk{\fs}(0)) \tpkt
  \end{equation*}
  Let $C = \cs_1(\hole,\hole)$, remember that $\repr{\cs_1} = \cns(\prs(x,\hole), \hole)$, 
  $\reprs{\hole} = \hole$
  and observe that $C' = \cns(\prs(0,\hole), \hole) \in \reprs{C}$
  by taking the substitution $\sigma = \set{x \mapsto 0}$.
  And hence we can reformulate the above two steps as
  $\mf(\ms(0)) \rew[\RSe] C'[\gs(0), \fs(0)]$ and likewise
  $\mrk{\mf}(\ms(0)) \rew[\TPWIDP(\RSe)] 
  C[\mrk{\gs}(0), \mrk{\fs}(0)]$.
\end{example}
We manifest the above observation in the following lemma.
\begin{lemma}\label{l:tpwidp:sizesimul}
  Let $s \in \Tb$ be a ground and basic term. Suppose $s \irss t$. 
  Let $\PS = \TPWIDP(\RS)$ and let $\US = \URs(\WIDP(\RS))$.
  Then there exists contexts $C'
  \in \CCtx$, $C \in \reprs{C'}$ and terms $\seq{t}$ such that 
  $t = C[\seq{t}]$ and moreover, $\mrk{s} \irss[\PS \cup \US] C'[\mrk{t_1},\dots,\mrk{t_n}]$.
\end{lemma}
\begin{proof}
  We proof the lemma by induction on the length of the rewrite sequence $s \irewpos{n} t$. 
  The base case $n=0$ is trivial, we set $C = C' = \hole$. 
  So suppose $s \irewpos{n} t \irew u$ and the property holds for $n$. 
  And thus we can find contexts $C'_t \in \CCtx$, $C_t \in \reprs{C'_t}$ and terms $\seq{t}$ such that 
  $t = C_t[\seq{t}]$ and moreover, $\mrk{s} \irss[\PS \cup \US] C'_t[\mrk{t_1},\dots,\mrk{t_n}]$.
  Without loss of generality we can assume $u = C_t[t_1,\dots,u_i,\dots,t_n]$ 
  with $t_i \irew u_i$, as the context $C_t$ is solely build
  from constructors and normal forms of $\RS$. 
  
  First, suppose $t_i \irewt u_i$, and hence $t_i = {l\sigma}$ for
  ${l \to r} \in \RS$ and substitution $\ofdom{\sigma}{\VS \to
    \NF(\RS) \cap \TA(\FS)}$. 
  Moreover ${\mrk{l} \to \cs(\mrk{r_1},\dots,\mrk{r_m})} \in \PS$ such that $u_i =
  (\repr{\cs}\sigma)[r_1\sigma,\dots,r_m\sigma]$. We set $C'$ as the
  context obtained from replacing the $i$-th hole of $C'_t$ by
  $\cs(\hole,\dots,\hole)$, likewise we set $C$ as the 
  context obtained from replacing the $i$-th hole of $C_t$ by
  $\repr{\cs}\sigma$. Note that $C \in \reprs{C'}$.
  We conclude 
  $\mrk{s} \irss[\PS \cup \US] C'[\mrk{t_1},\dots,{\mrk{r_1}\sigma,\dots,\mrk{r_m}\sigma},\dots,\mrk{t_n}]$
  and $u = C[t_1,\dots,r_1\sigma,\dots,r_m\sigma,\dots,t_n]$ which
  establishes the lemma for this case. 
  
  Now suppose $t_i \irew u_i$ is a step below the root. Thus we have 
  also $\mrk{t_i} \irew \mrk{u_i}$. As shown in
  \cite[Lemma 16]{HirokawaMoser:2008}, the latter can be strengthened to 
  $\mrk{t_i} \qrew[\US]{\PS \cup \US} \mrk{u_i}$.
  We conclude $\mrk{s} \irss[\PS \cup \US]
  C'_t[\mrk{t_1},\dots,\mrk{u_i},\dots,\mrk{t_n}]$, 
  and the lemma follows by setting $C' = C'_t$ and $C = C_t$. 
\end{proof}  

Suppose $\WIDP(\RS)$ contains non-nullary
compound symbols.
In order to establish an embedding in the sense of Lemma~\ref{l:embed} for that case, 
by the above lemma we see that it suffices to 
consider only terms of shape $s = C[{\mrk{s_1},\dots,\mrk{s_n}}]$
with $C \in \CCtx$.
With this insight, we adjust Lemma~\ref{l:embed} as below.
Observe that due to the definition of $\intq$, we 
cannot simply apply Lemma~\ref{l:embed} together with 
closure under context of $\gpopv_k$ here.
\begin{lemma}\label{l:intq:embed}
  Let $s = C[\mrk{s_1},\dots,\mrk{s_n}]$ for $C \in \CCtx$ and
  $\seq{s} \in \TERMS$.
  Let $\PS = \TPWIDP(\RS)$ and $\US = \URs(\WIDP(\RS))$.
  There exists a uniform constant $k\in \NAT$ depending only on $\RS$
  such that if $\PS \subseteq {\gpopp}$ holds then $s \vrew[\PS] t$
  implies $\intq(s) \gpopv_k \intq(t)$.
  Moreover, if ${\US} \subseteq {\geqpopp}$ holds then 
  $s \vrew[\US] t$ implies $\intq(s) \geqpopv_k \intq(t)$.
\end{lemma}
\begin{proof}
  We proof the lemma for 
  $k \defi \max\set{3 \cdot \bN{r} \mid {l \to r} \in \PS \cup \US}$.
  Suppose $s \vrew[\PS] t$ or $s \vrew[\US] t$ respectively, and thus 
  $t = C[\mrk{s_1},\dots,t_i,\dots,\mrk{s_n}]$ for some term $t_i$.
  There exists a context $C'$ (over sequences) such that
  $\intq(s) = C'[\intq(\mrk{s_i})]$ and 
  $\intq(t) = C'[\intq(\mrk{t_i})]$.
  First assume $\mrk{s_i} \vrew[\PS] t_i$, and thus 
  $\intq(\mrk{s_i}) = [\intn(\mrk{l}\sigma)]$ and 
  $\intq(t_i) = [[\intn(\mrk{r_1}\sigma)],\dots,[\intn(\mrk{r_m}\sigma)]]$
  for ${l \to \cs(\mrk{r_1},\dots,\mrk{r_m})} \in \PS$. 
  To verify $\intq(s) \gpopv_k \intq(t)$,
  by Definition~\ref{d:approx}(\ref{d:approx:ii}) 
  and Definition~\ref{d:approx}(\ref{d:approx:iv}),
  it suffices to verify
  $\intn(\mrk{l}\sigma) \gpopv_{k-1} \intn(\mrk{r_j}\sigma)$ for all $j \in \set{1,\dots,m}$. 
  The latter is an easy consequence of Lemma~\ref{l:embed}, where 
  we employ that (i) $\mrk{l} \gpopp \mrk{r_j}$ follows from the assumption $\PS \subseteq {\gpopp}$, 
  and (ii) $\bN{\pi(r)} > \bN{\pi(r_j)}$. Both properties are straight
  forward to verify since $\pi$ is safe.
  For $\mrk{s_i} \vrew[\US] t$ we have 
  $\intq(\mrk{s_i}) = [\intn(\mrk{s_i})]$ and 
  $\intq(\mrk{t_i}) = [\intn(\mrk{t_i})]$ for ${l \to r} \in \US$.
  From Lemma~\ref{l:embed} we obtain $\intn(\mrk{s_i}) \geqpopv_k \intn(\mrk{t_i})$ which establishes the lemma.
\end{proof}

The proof of Theorem~\ref{t:widp} is now easily obtained by
incorporating the above lemma into Theorem~\ref{t:relstep:root}.
\begin{theoremwidp}
    Let $\RS$ be a constructor TRS, and let $\PS$ denote the set of weak innermost 
    dependency pairs. 
    Assume $\PS$ is non-duplicating, and suppose ${\URs(\PS)} \subseteq {>_\AS}$ 
    for some SLI $\AS$. Let $\pi$ be a safe
    argument filtering.
    If $\PS \subseteq {\gpopp}$ and $\URs(\PS) \subseteq {\geqpopp}$ then
    $\rcRi$ is polynomially bounded. 
\end{theoremwidp}

\begin{proof}
  According to Proposition~\ref{p:dp} we need to find a polynomial $p$ 
  such that 
  \begin{equation*}
    \dl(\mrk{t},\irrew{\WIDP(\RS)}{\URs(\WIDP(\RS))}) \leqslant p(\size{\mrk{t}}) \tpkt
  \end{equation*}
  We set $\PS = \TPWIDP(\RS)$ and likewise $\US = \URs(\WIDP(\RS))$. 
  Clearly, it suffices to show $\dl(\mrk{t},\irrew{\TPWIDP(\RS)}{\URs(\WIDP(\RS))})
  \leqslant p(\size{\mrk{t}})$ for that.
  Consider a sequence
  $$
  \mrk{t} = t_0
  \irrew{\PS}{\US} t_1
  \irrew{\PS}{\US} \dots
  \irrew{\PS}{\US} t_\ell\tkom
  $$
  and pick a relative step $t_i \irrew{\PS}{\US} t_{i+1}$. 
    Define $\US' = \US \cup \VS(\PS \cup \US)$ and $\nfV{t} = \nfV[\PS
  \cup \US]{t}$. 
  Clearly Lemma~\ref{l:v:simulation:relative} can be extended
  to account for steps of $\PS$ below the root, and thus
  $\nfV{t_i} \vrrew{\PS}{\US'} \nfV{t_{i+1}}$ follows. 
  Hence for some terms $u$ and $v$, 
  $\nfV{t_i} \vrss[\US'] u \vrew[\PS] v \vrss[\US'] \nfV{t_{i+1}}$.
  As shown in Lemma~\ref{l:tpwidp:sizesimul}, 
  all involved terms in the above sequence have the shape
  $C[\mrk{s_1},\dots,\mrk{s_n}]$, $C \in \CCtx$.
  As $\WIDP(\RS) \subseteq {\gpopp}$, and since $\pi$ is safe, it is
  easy to infer that $\PS \subseteq {\gpopp}$ holds (we just set every
  compound symbol from $\PS$ minimal in the precedence). 
  And hence Lemma~\ref{l:intq:embed} translates the above relative step to
  $\intq(\nfV{s}) \gpopv^+_k \intq(\nfV{t})$ for 
  some uniform constant $k$.
  As a consequence, $\dl(t,\irrew{\WIDP(\RS)}{\URs(\WIDP(\RS))}) 
  \leqslant \Slow(\intq(\nfV{t}))$ for all terms $t$. 
  Fix some reducible and basic term $t \in \Tb$.
  Observe $\intq(\nfV{\mrk{t}}) = [\intn(\mrk{t})]$ and so 
  from Lemma~\ref{d:pred:int} we see that 
  $\Slow(\intq(\nfV{\mrk{t}}))$ is bounded polynomially 
  in the size of $t$. The polynomial depends only on $k$. We conclude the theorem.
\end{proof}

\subsection{Proof of Theorem~\ref{t:polytime}} \label{s:proof:polytime}
We now proceed with the proof Theorem~\ref{t:polytime}, 
which is essentially an extension to Theorem~\ref{t:widp}. 

We first precisely state what it means that a TRS \emph{computes} 
some function. For this,
let $\ofdom{\enc{\cdot}}{\Sigma^\ast \to \TA(\CS)}$ 
denote an \emph{encoding function} that represents words over the
alphabet $\Sigma$ as ground values. 
We call an encoding $\enc{\cdot}$ \emph{reasonable} if it is bijective and 
there exists a constant $c$ such that 
$\size{u} \leqslant \size{\enc{u}} \leqslant c \cdot \size{u}$ for
every $u \in \Sigma^*$. 
Let $\enc{\cdot}$ denote a reasonable encoding function, and let
$\RS$ be a completely defined, orthogonal and terminating TRS.
We say that an $n$-ary function $f \colon (\Sigma^\ast)^n \to \Sigma^*$
is \emph{computable} by $\RS$ if there exists a defined function symbol $\mf$ 
such that for all $w_1,\dots,w_n,v \in \Sigma^\ast$
$\mf(\enc{w_1},\dots,\enc{w_n}) \to^! \enc{v} \Longleftrightarrow f(w_1,\dots,w_n) = v$.
On the other hand the TRS $\RS$ \emph{computes} $f$, 
if the function $f \colon (\Sigma^\ast)^n \to \Sigma^*$
is defined by the above equation. 

Below we abbreviate $\Q_\pi$ as $\Q$ for
predicative interpretation $\Q \in \set{\ints[],\intn[],\intq[]}$
and the particular argument filtering $\pi$ that 
induces the identity function on terms.
Consider the following lemma.

\begin{lemma}\label{l:tpwidp:sizebound}
  Let $\RS$ be an $S$-sorted and completely defined constructor TRS
  such that the underlying signature is simple. 
  If ${\TPWIDP(\RS) \cup \URs(\WIDP(\RS))} \subseteq {\geqpop}$ then 
  there exists a polynomial $p$ such that for all ground and well-typed basic terms
  $t \in \Tb$, $\mrk{t} \irss[\TPWIDP(\RS) \cup \URs(\WIDP(\RS))] s$ implies
  $\size{s} \leqslant p(\size{t})$. 
\end{lemma}
\begin{proof}
  Let $\RSS = \TPWIDP(\RS) \cup \URs(\WIDP(\RS))$. 
  Suppose $\mrk{t} \irss[\RSS] s$, or equivalently $\mrk{t} \vrss[\RSS] s$ since $\RS$ is completely defined.
  By Lemma~\ref{l:intq:embed} we derive $\intq[](\mrk{t}) \geqpopv_k^* \intq[](s)$ 
  for some uniform $k \in \NAT$.
  And thus $\Slow(\intq[](s)) \leqslant \Slow(\intq[](\mrk{t}))$.
  As $\Slow(\intq[](\mrk{t})) = \Slow([\intn[](\mrk{t})])$ 
  is bounded polynomially in the size of $t$ according to Lemma~\ref{lem:slow:intn}, we
  see that there exists a polynomial $p$ such that $\Slow(\intq[](s))
  \leqslant \Slow(\intq[](\mrk{t})) \leqslant p(\size{t})$.
  Since $\RS$ is and $S$-sorted TRS over a simple signature, 
  the same holds for $\RSS$ due to Lemma~\ref{l:tpwidp:preserve}.
  And thus since $\mrk{t}$ is well-typed and $\mrk{t} \irss[\RSS] s$ holds, also $s$ is well-typed. 
  Let $q$ be the polynomial as given from Lemma~\ref{l:slow:sizebound} with 
  $ \size{s} \leqslant q(\Slow(\intq[](s)))$.
  Summing up, we derive 
  $ \size{s} \leqslant q(\Slow(\intq[](s))) \leqslant q(p(\size{t}))$ 
  as desired. 
\end{proof}

The above lemma has established that sizes of reducts with respect to
the relation
$\irew[\TPWIDP(\RS) \cup \URs(\WIDP(\RS))]$ are bounded polynomially
in the size of the start term, provided we can orient dependency pairs
and usable rules. It remains to verify that this is indeed sufficient to appropriately estimate sizes of reducts with respect to $\irew$. 
The fact is established in the final Theorem.
\begin{theorempoly}
  Let $\RS$ be an orthogonal $S$-sorted and completely defined
  constructor TRS such that the underlying signature is simple.
  Let $\PS$ denote the set of weak innermost 
  dependency pairs. 
  Assume $\PS$ is non-duplicating, and suppose ${\URs(\PS)} \subseteq {>_\AS}$ 
  for some SLI $\AS$.
  If $\PS \subseteq {\gpopp}$ and $\URs(\PS) \subseteq {\geqpopp}$ then
  the functions computed by $\RS$ are computable in polynomial-time.
\end{theorempoly}
\begin{proof}
  We single out one of the defined symbols $\mf \in \DS$ and consider the corresponding
  function $f \colon (\Sigma^{\ast})^n \to \Sigma^{\ast}$ computed by $\RS$. 
  Under the assumptions, $\RS$ is terminating, but moreover
  $\rcRi$ is polynomially bounded according to Theorem~\ref{t:widp}. 
  Additionally, from orthogonality (and hence confluence) of $\RS$, 
  normal forms are unique and so the function
  $f$ is well-defined. 
  Suppose $\mf(\enc{w_1},\dots,\enc{w_n}) \rsn \enc{v}$ for words $w_1,\dots,w_n,v$. 
  In particular, from confluence we see that 
  \begin{equation*}
    \mf(\enc{w_1},\dots,\enc{w_n}) \irew t_1 \irew \cdots \irew t_{\ell} = \enc{v}\tpkt
  \end{equation*}
  It is folklore that there exists a polytime algorithm performing one
  rewrite step. Hence to conclude the existence of a polytime algorithm for $f$ it suffices to 
  bound the size of terms $t_i$ for $1 \leqslant i \leqslant \ell$ polynomially
  in $\sum_i \size{w_i}$. And as we suppose that the encoding $\enc{\cdot}$ is reasonable, 
  it thus suffice to bound the sizes of $t_i$ for $i \in \set{1,\dots,\ell}$ 
  polynomially in the size of $t_0 = \mf(\enc{w_1},\dots,\enc{w_n})$. 

  Consider a term $t_i$. Without loss of generality, we can assume $t_i$ is ground. 
  According to Lemma~\ref{l:tpwidp:sizesimul}
  there exists contexts $C'_i \in \CCtx$, $C_i \in \reprs{C'_i}$ and terms $\seq{u}$ such that 
  $t_i = C_i[\seq{u}]$ and moreover, $\mrk{t_0} \irss[\PS \cup \US]
  C'_i[\mrk{u_1},\dots,\mrk{u_n}]$ for all $i \in \set{1,\dots,\ell}$.
  From the assumption $\WIDP(\RS) \subseteq {\gpop}$ we see 
   $\TPWIDP(\RS) \subseteq {\gpop}$.
  Thus by Lemma~\ref{l:tpwidp:sizebound} there exists a polynomial $p$ such that 
  $\size{C'_i[\mrk{u_1},\dots,\mrk{u_n}]} \leqslant p(\size{t_0})$.
  And so, clearly $\sum_{j=0}^n \size{u_j} \leqslant p(\size{t_0})$.
  It remains to bound the sizes of contexts $C_i$ polynomially in $\size{t_0}$.

  Recall Definition~\ref{d:reprs}, and recall that $C_i \in \reprs{C'_i}$.
  Thus $C_i$ is a context build from constructors and variables,
  where the latter are replaced by normal forms of $\RS$.
  Since $\RS$ is completely defined, $\NF(\RS)$ coincides with
  values. We conclude that $C_i \in \TA(\CS \cup \set{\hole_s \mid s
    \in S})$.
  Here $\hole_s$ denotes the hole of sort $s$. 
  Moreover since $\RS$ is $S$-sorted, and $t_0 \irss C_i[\seq{u}]$, we
  see that $C_i$ is well-typed.
  We define $\dlt = \max\set{\depth(r) \mid {l \to r} \in \RS}$. 
  By a straight forward induction it follows that 
  $\depth(t_i) \leqslant \depth(t_0) + \dlt \cdot i \leqslant \size{t_0} + \dlt \cdot \dl(t_0,\irew)$.
  As a consequence, $\depth({C_i}) \leqslant \size{t_0} + \dlt \cdot
  \dl(t_0,\irew)$, and thus by Proposition~\ref{prop:ss} there exists constants $c,d\in\NAT$ such that 
  $\size{C_i} \leqslant c \cdot \depth(C_i)^d \leqslant c \cdot (\size{t_0} + \dlt \cdot \dl(t_0,\irew))^d$. 
  As we have that $\dl(t_0,\irew)$ is polynomially 
  bounded in the size of $t_0$, it follows that $\size{C_i}\leqslant q(\size{t_0})$ for some
  polynomial $q$.

  Summing up, we conclude that for all $i \in \set{1,\dots,\ell}$, 
  $\size{t_i} \leqslant p(\size{t_0}) + q(\size{t_0})$ for the
  polynomials $p$ and $q$ from above. 
  This concludes the theorem. 
\end{proof}

\end{document}